\documentclass[preprint,review,12pt]{elsarticle}

\usepackage{graphics}

\usepackage{amssymb}
\usepackage{subeqn}

\usepackage{epstopdf}
\usepackage{subfigure}
\usepackage{makecell}

\biboptions{sort&compress}

\journal{}

\begin{document}

\begin{frontmatter}


\title{Axisymmetric lattice Boltzmann model for multiphase flows with large density ratio}


\author[a]{Hong Liang}
\author[b]{Yang Li}
\author[a]{Jiangxing Chen}
\author{Jiangrong Xu\fnref{a}\corref{cor1}}\cortext[cor1]{Corresponding author.
\hspace*{20pt}}\ead{lianghongstefanie@163.com (Hong Liang); jrxu@hdu.edu.cn}

\address[a]{Department of Physics, Hangzhou Dianzi University, Hangzhou, 310018, China}
\address[b]{Department of Mathematics, Hangzhou Dianzi University, Hangzhou, 310018, China}

\begin{abstract}

In this paper, a novel lattice Boltzmann (LB) model based on the
Allen-Cahn phase-field theory is proposed for simulating
axisymmetric multiphase flows. The most striking feature of the
model is that it enables to handle multiphase flows with large
density ratio, which are unavailable in all previous axisymmetric LB
models. The present model utilizes two LB evolution equations, one
of which is used to solve fluid interface, and another is adopted to
solve hydrodynamic properties. To simulate axisymmetric multiphase
flows effectively, the appropriate source term and equilibrium
distribution function are introduced into the LB equation for
interface tracking, and simultaneously, a simple and efficient
forcing distribution function is also delicately designed in the LB
equation for hydrodynamic properties. Unlike many existing LB
models, the source and forcing terms of the model arising from the
axisymmetric effect include no additional gradients, and
consequently, the present model contains only one non-local phase
field variable, which in this regard is much simpler. In addition,
to enhance the model's numerical stability, an advanced
multiple-relaxation-time (MRT) model is also applied for the
collision operator. We further conducted the Chapman-Enskog analysis
to demonstrate the consistencies of our present MRT-LB model with
the axisymmetric Allen-Cahn equation and hydrodynamic equations. A
series of numerical examples, including static droplet, oscillation
of a viscous droplet, breakup of a liquid thread, and bubble rising
in a continuous phase, are used to test the performance of the
proposed model. It is found that the present model can generate
relatively small spurious velocities and can capture interfacial
dynamics with higher accuracy than the previously improved
axisymmetric LB model. Besides, it is also found that our present
numerical results show excellent agreement with analytical solutions
or available experimental data for a wide range of density
ratios, which highlights the strengths of the proposed
model.

\end{abstract}


\end{frontmatter}



\section{Introduction}

Multiphase fluid flows are ubiquitous in nature and are of
considerable interest in both scientific and engineering fields. In
spite of improving experimental studies of these multiphase
phenomena, numerical modeling becomes an increasingly important
approach with the rapid development of computer technology and
gradual enrichment of computing methods. And also, it can provide a
convenient access to physical quantities, such as variational
interface shapes, fluid velocity across interface, pressure
distribution inside and outside bulk phase, which are usually
difficult to measure experimentally. Nonetheless, it still remains
an intractable task to the modeling of multiphase flows and further
develop efficient numerical algorithms that can accurately describe
physical phenomena behind such flows. The reasons behind the
challenges lie in the complexity of interface dynamics among
multispecies fluids, density and viscosity jumps across the
interface, and surface tension force modeling. Several numerical
methods to date have been proposed for simulating multiphase flows,
which can be roughly divided into two categories: sharp-interface
methods, and diffusion-interface methods. The former methods
commonly include the volume-of-fluid method~\cite{Hirt},
front-tracking method~\cite{Unverdi} and level set
method~\cite{Sussman}. In this type of traditional methods, one
might solve continuum mechanics equations coupled with a suitable
technique to track the phase interface. The interface needs to be
captured manually based on some complex ad-hoc criteria, and fluid
properties in these methods such as density and viscosity vary
sharply at the interface. Therefore, they in this regard are not
suitable for handling multiphase flows with large interface
topological change~\cite{Anderson, Ding}. As for the latter methods,
one replaces the sharp interface with a transition region across
which fluid physical properties are allowed to change smoothly. This
feature makes them more potential for simulating complex interfacial
dynamic problems. Among diffusion-interface approaches, the
phase-field method~\cite{Jacqmin} and lattice Boltzmann (LB)
method~\cite{Guo, Kruger} are two popular ones. In the phase-field
method, the thermodynamic behavior of a multiphase system is
described by the free energy as a function of an order parameter,
which is used to capture phase interface. The interfacial governing
equation for the order parameter is formulated as the convective
Cahn-Hilliard~\cite{Cahn, Jacqmin} or Allen-Cahn
(AC)~\cite{Sun,Chiu} equation, which is usually solved by using a
finite difference like scheme in phase field modelling. Therefore
the phase field method will inherit some certain weaknesses rooted
in the conventional numerical methods.

Alternatively, the LB method~\cite{Guo, Kruger} has received great
attention for modelling hydrodynamic phenomena and in particular for
multiphase flows. It is a mesoscopic method based on the kinetic
equation for the particle distribution function, connecting the
bridge between the macroscopic continuum model and molecular
dynamics. Due to its mesoscopic nature, the LB method has several
distinct advantages over the traditional numerical methods, such as
the simplicity of algorithmic, nature parallelization and easy
implementation of complex boundary. Particularly, the intermolecular
interactions in a multiphase system can be incorporated directly in
the framework of LB method, while they are difficult to handle in
traditional numerical methods. Historically, from different physical
pictures of the interactions, several types of LB models have been
established for simulating multiphase flows, which include the
color-gradient model~\cite{Gunstensen}, pseudo-potential
model~\cite{Shan}, free energy model~\cite{Swift}, and phase-field
based model~\cite{He, Lee, Liang, Liang1}. Some advanced LB models
based on these original models have also been proposed in succession
and interesting readers can refer to good reviews~\cite{QLi, Liu} of
this field and the references therein.

In practice, there exists many multiphase fluid problems that
display axial symmetry. Examples include head-on collision of binary
droplets~\cite{Nobari, PremnathKN}, bubble rising in a continuous
phase~\cite{Hua, Huang}, droplet formation in
micro-channel~\cite{Link, Utada}, droplet splashing on a solid
surface or wetting liquid film~\cite{Bussmann, Josseranda}, and so
on. The most natural manner to simulate such axisymmetric flows is
to apply a three-dimensional (3D) LB multiphase model with suitable
curved boundary conditions. This treatment, however, does not take
any advantages of the axisymmetric property of flow. If we recognize
the fact that 3D axisymmetric flows can reduce to the
two-dimensional (2D) ones in meridian plane, an effective approach
in this regard is to develop quasi-2D LB models for simulating these
flows. Up to now, some scholars have made an effort to construct
effective axisymmetric models based on the LB method, whereas most
of them are proposed for single-phase flows within the isothermal
and thermal systems~\cite{Halliday, Peng, Guo1, Li, ZhengL, Zhang}. The
first axisymmetric LB model for multiphase flows was attributed to
Premnath and Abraham~\cite{Premnath}, who introduced some suitable
source terms in the Cartesian coordinate multiphase model of He
{\it{et al.}}~\cite{He}. The source terms are used to account for
the axisymmetric contribution of inertial, viscous and surface
tension forces, while they contains many complex gradients of
density and velocity and thus undermine the simplicity of the model.
In addition, the model has a drawback inherited in He's model that
the available highest density ratio is limited to 15. And also based
on the multiphase model of He {\it{et al.}}~\cite{He}, Huang {\it{et
al.}}~\cite{Huang} presented an improved axisymmetric LB model, in
which a mass correction step is imposed in every numerical
iteration. They applied this model to the simulation of bubble
rising, and found rather good agreement with experiments. However,
the highest density ratio in the simulation is limited to 15.5, and
the model would undergo numerical instability with the increasing
density ratio, as they stated. To remove this limitation, Mukherjee
and Abraham \cite{Mukherjee} proposed an axisymmetric counterpart
based on the high-density-ratio model of Lee and Lin~\cite{Lee}, in
which some source terms relating to the density and velocity
gradients are also included. To improve numerical stability, they
also used a three-stage mixing discretization scheme, which
unfortunately could induce the violation of the mass
conservation~\cite{Lou}. Later, Huang {\it{et al.}}~\cite{HuangJJ}
put forward a hybrid LB model for axisymmetric binary flows, where a
finite difference scheme is used to solve the convective
Cahn-Hilliard equation for interface capturing and a
multiple-relaxation-time (MRT) LB scheme is adopted for solving the
Navier-Stokes (NS) equations. The densities of binary fluids are
supposed to be uniform in the given NS equations, and therefore
their model in theory can only be applicable for density-matched
two-phase flows. The extension of the popular pseudo-potential
model~\cite{Shan} to the axisymmetric version was conducted by
Srivastava {\it{et al.}}~\cite{Srivastava}. Recently, Liang {\it{et
al.}}~\cite{Liang2} proposed a Cahn-Hilliard phase-field based LB
model for axisymmetric multiphase flows. One distinct feature of
this model is that the added source terms representing the
axisymmetric effect contain no gradients, thus simplifying the
computation. The model is also demonstrated to be accurate for
simulating multiphase flows with moderate density ratios, while it
is unable to tackle large-density-ratio cases. More recently,
Liu~{\it{et al.}}~\cite{LiuH} developed a color-gradient LB model
for axisymmetric multicomponent flows. This model can deal with
binary fluids with high viscosity ratio, whereas the density ratio
considered is very small.

As reviewed above, several types of LB models have been proposed for
axisymmetric multiphase flows, and all these models are restricted
to the flows with small or moderate density contrasts. Generally,
the density ratio can approach 1000 for a realistic liquid-vapor
system, and to develop a high-density-ratio multiphase model is very
attractive in LB community. In this paper, we intend to present a
simple, accurate and also robust LB model for axisymmetric
multiphase flows, which can tolerate large density contrasts. The
proposed LB model is based on the AC phase field theory, which
involves a lower-order diffusion term compared with the
Cahn-Hilliard equation, and thus is expected to achieve a better
numerical accuracy. Modified equilibrium distribution functions and
simplified forcing distribution functions are also incorporated in
the model to recover the correct axisymmetric AC and NS equations.
Besides, unlike most of available LB models~\cite{Premnath, Huang,
Mukherjee, HuangJJ, Srivastava}, the introduced source terms arising
from the axisymmetric effect contain no gradients in the present
model. The rest of this paper is organized as follows. In Sec
\ref{sec:method}, the governing equations and their axisymmetric
formulations are first given, and then a novel axisymmetric LB model
based on the AC phase-field theory are proposed. Numerical
validations for the present model can be found in Sec
\ref{sec:result}, and at last, we made a brief summary in Sec.
\ref{sec:sum}.

\section{GOVERNING EQUATIONS AND MATHEMATICAL MODEL}\label{sec:method}

\subsection{Governing equations}
The AC equation contains only at most a second-order gradient
diffusion term, which can be more efficient and less dispersive in
solving the phase interface compared with the commonly used
Cahn-Hilliard theory~\cite{WangHL, Fakhari, Chai, Liang3}. Therefore the
interface tracking equation in this study is built upon the AC phase
field theory. The original AC equation was not globally
conservative, and recently reformulated into the conservative
form~\cite{Chiu} based on the work of Sun and Beckermann~\cite{Sun}.
This specific formulation is commonly referred as the conservative
phase-field model and will be adopted here. Then the conservative AC
equation for interface tracking can be written as~\cite{Chiu,
WangHL, RenF},
\begin{equation}
{{\partial \phi}\over{\partial t}}+\nabla\cdot(\phi
\textbf{u})=\nabla \cdot {[M (\nabla {\phi}-\lambda \textbf{n})]},
\end{equation}
where $\phi$ is the order parameter acting as a phase field to
distinguish different fluids, $\textbf{u}$ is the fluid velocity,
$M$ is the mobility, $\textbf{n}$ is the unit vector normal to the
interface and is calculated as
$\textbf{n}={{\nabla\phi}/{|\nabla\phi|}}$, $\lambda$ can be
expressed by
\begin{equation}
\lambda=-{4(\phi-1)\phi\over{W}},
\end{equation}
where $W$ is the interfacial thickness, the phase field $\phi$ takes
1 and 0 in the bulk regions of the liquid and vapor fluids,
respectively, and $\phi=0.5$ indicates the phase interface of binary
fluids. To simulate hydrodynamic flows, the AC equation should be
coupled with the incompressible NS equations, which can be written
as~\cite{Unverdi, Liang3}
\begin{subequations}
\begin{equation}
\nabla  \cdot {\bf{u}} = 0,
\end{equation}
\begin{equation}
{{\partial ({\rho \bf{u}}}) \over {\partial t}} + \nabla \cdot (\rho
{\bf{u}}{\bf{u}}) = -\nabla p + \nabla  \cdot \left[ {\nu\rho(\nabla
{\bf{u}} + \nabla {{\bf{u}}^T})} \right] + {{\bf{F}}_s} + {\bf{G}},
\end{equation}
\end{subequations}
where $p$ is the hydrodynamic pressure, $\nu$ is the kinematic
viscosity, $\textbf{G}$ is the possible external force,
$\textbf{F}_s$ is the surface tension force. According to
Ref.~\cite{Kim}, there exists several treatments in terms of surface
tension force, which could give rise to different performances. To
reduce the spurious velocity, in this work we choose the widely used
the potential form $\textbf{F}_s=\mu\nabla\phi$~\cite{Liang,
Fakhari, Liang2}, where $\mu$ is the chemical potential given by
\begin{equation}
\mu=4\beta \phi(\phi-1)(\phi-0.5)-k\nabla^2\phi,
\end{equation}
where the parameters $\beta$ and $k$ can be determined by the
surface tension $\sigma$ and the interfacial thickness, i.e.,
$k=1.5\sigma W$, $\beta=12\sigma/W$.

We now performed the coordinate transformation to derive the
governing equations of isothermal multiphase flow in the
axisymmetric system. The transformation is given by
\begin{center}
$(x,y,z)\rightarrow(r,\theta,z)$
\end{center}
with the relations $x=r\cos\theta$, $y=r\sin\theta$, where $r$, $z$,
$\theta$ denote the radial, axial and azimuthal directions,
respectively. The flows are assumed to have no swirl here and thus
we can set the azimuthal velocity and azimuthal coordinate
derivatives to be zero. In this case, the resulting macroscopic
equations in the axisymmetric framework can be expressed by
\begin{equation}
\partial_t\phi+\partial_\alpha(\phi u_\alpha)+\frac{\phi{u_r}}{r}=\partial_\alpha(M\partial_\alpha \phi)+\frac{M\partial_r
\phi}{r}-M\partial_\alpha(\lambda{n_\alpha})-\frac{M\lambda{\partial_r}\phi}{r|\nabla\phi|}
\end{equation}
and
\begin{subequations}
\begin{equation}
\partial_\alpha u_\alpha+\frac{u_r}{r}=0,
\end{equation}
\begin{equation}
\partial_t (\rho {u_\beta})+{\partial_\alpha}(\rho
{u_\alpha}u_\beta)=-\partial_\beta{p}+{\partial_\alpha}\nu\rho({\partial_\alpha}u_\beta+{\partial_\beta}u_\alpha)
+{\tilde{F}_{s\beta}}+G_\beta+\frac{\nu\rho(\partial_r{u_\beta}+\partial_\beta{u_r})}{r}-\frac{2\rho\nu{u_r}\delta_{{\beta}r}}{r^2}-{{\rho u_r u_\alpha} \over {r}},
\end{equation}
\end{subequations}
where $\alpha,\beta=\{r,z\}$, $\delta$ is the Kronecker function,
${\tilde{F}_{s\beta}}$ is the modified surface tension force given
by ${\tilde{F}_{s\beta}}=\tilde{\mu}\nabla{\phi}$, and
$\tilde{\mu}=\mu-k\partial_r{\phi}/r$. From Eqs. (5) and (6), it can
be clearly found that some additional source terms are generated due
to the axisymmetric effect. These source terms contain some
gradients on the fluid velocity, and seem to be implemented
complicatedly. To our knowledge, most of previously proposed
axisymmetric LB models~\cite{Halliday, Li, Zhang, Premnath, Huang,
Mukherjee, HuangJJ, Srivastava} for hydrodynamic flows were
constructed based on the governing equation (6), which thus makes
them more complex than the standard models in the Cartesian
coordinate system. Alternatively, we convert the governing equations
(5) and (6) in another form. With some algebraic manipulations, the
derived macroscopic equations can be presented as
\begin{equation}
\partial_t(r\phi)+\partial_\alpha(r\phi u_\alpha+M\phi \delta_{\alpha r})=\partial_\alpha[M\partial_\alpha (r\phi)-M r\lambda
n_\alpha]
\end{equation}
and
\begin{subequations}
\begin{equation}
\partial_\alpha(r u_\alpha)=0,
\end{equation}
\begin{equation}
\partial_t (r\rho {u_\beta})+{\partial_\alpha}(r\rho
{u_\alpha}u_\beta)=-\partial_\beta(rp)+{\partial_\alpha}[r\nu\rho({\partial_\alpha}u_\beta+{\partial_\beta}u_\alpha)]+r(\tilde{F}_{s\beta}+G_{\beta})+(p-{2\rho\nu
\over r}u_r)\delta_{\beta r}.
\end{equation}
\end{subequations}
The present LB model will be built upon the macroscopic equations
(7) and (8), which utilizes two LB evolution equations, one of them
is used to solve the axisymmetric AC equation, and another for
solving the axisymmetric NS equations. It will be demonstrated below
that the introduced source terms representing the axisymmetric
contributions contain no gradients in the present model.

\subsection{Axisymmetric LB model for the Allen-Cahn equation}
Based on the collision operators, the LB approach can be roughly
divided into four categories, including the MRT model~\cite{Luo1},
the two-relaxation-time model~\cite{Ginzburg}, the
single-relaxation-time model~\cite{Qian}, and the entropic LB
model~\cite{Karlin}. To date, these models all have their own
impressive versatility in simulating hydrodynamic flows, while the MRT
model in comparison has its superiority in terms of stability and accuracy, and thus will be used in the
present LB modelling for multiphase flows. The LB evolution equation
of the MRT model for the axisymmetric AC equation can be proposed as
\begin{equation}
f_i(\textbf{x}+\textbf{c}_i\delta_t,t+\delta_t)-f_i(\textbf{x},t)=\Omega_i(\textbf{x},t)+\delta_t{F}_i(\textbf{x},t),
\end{equation}
where the collision operator $\Omega_i$ is defined by
\begin{equation}
\Omega_i(\textbf{x},t)=-({{{\bf{M}}^{-1}}{{\bf{S}}^f}{\bf{M}})_{ij}}[f_j(\textbf{x},t)-f_j^{eq}(\textbf{x},t)],
\end{equation}
where $f_i(\textbf{x},t)$ is the phase-field distribution function
associated with the discrete velocity $\textbf{c}_i$ at position
$\textbf{x}$ and time $t$, $f_i^{eq}(\textbf{x},t)$ is the
equilibrium distribution function. To match the target equation, we
design a new equilibrium distribution function as
\begin{equation}
f_i^{eq}=\omega_i\left[ r\phi+{c_{i\alpha}(r\phi
u_\alpha+M\phi\delta_{\alpha r}) \over {c_s^2}} \right],
\end{equation}
where $c_s$ is lattice sound speed, $\omega_i$ are the weighting
coefficients. $\omega_i$ and $\textbf{c}_i$ depend on the choice of
the discrete-velocity lattice model. For plane flows, the popular
D2Q9 lattice model~\cite{Qian, Liang, Liang1, Wei} is adopted here,
and $\omega_i$ can be then given by $\omega_0=4/9$,
$\omega_{1-4}=1/9$, $\omega_{5-8}=1/36$, and $\textbf{c}_i$ is
defined as
\begin{equation}
\mathbf{c}_{i}=\left\{
\begin{array}{ll}
 (0,0)c,                                                         & \textrm{ $i=0$},   \\
 (\cos [(i-1)\pi /2],\sin [(i-1)\pi /2])c,                       & \textrm{ $i=1-4$}, \\
 \sqrt{2}(\cos [(i-5)\pi /2+\pi /4],\sin [(i-5)\pi /2+\pi /4])c, & \textrm{ $i=5-8$},
\end{array}
\right.
\end{equation}
where $c=\delta_x/\delta_t$ denotes the lattice speed with
$\delta_x$ and $\delta_t$ representing the lattice spacing and the
time step, respectively, and $c_s=c/\sqrt{3}$. By convention,
$\delta_x$ and $\delta_t$ in the study of multiphase flows are set
to be length and time units, i.e., $\delta_x=\delta_t=1$. In Eq.
(10), $\textbf{M}$ is the transformation matrix, which is used to
project the distribution function in the discrete-velocity space
onto the ones in the moment space. Based on the D2Q9 lattice
structure, it can be given by~\cite{Luo1}
\begin{displaymath}
\bf{M}= \left(
\begin{array}{rrrrrrrrr}
  1 & 1 & 1 & 1 & 1 & 1 & 1 & 1 & 1 \\
  -4 & -1 & -1 & -1 & -1 & 2 & 2 & 2 & 2 \\
  4 & -2 & -2 & -2 & -2 & 1 & 1 & 1 & 1 \\
  0 & 1 & 0 & -1 & 0 & 1 & -1 & -1 & 1 \\
  0 & -2 & 0 & 2 & 0 & 1 & -1 & -1 & 1 \\
  0 & 0 & 1 & 0 & -1 & 1 & 1 & -1 & -1 \\
  0 & 0 & -2 & 0 & 2 & 1 & 1 & -1 & -1 \\
  0 & 1 & -1 & 1 & -1 & 0 & 0 & 0 & 0 \\
  0 & 0 & 0 & 0 & 0 & 1 & -1 & 1 & -1
\end{array}\right),
\end{displaymath}
and ${\bf{S}}^f$ in Eq. (10) is a diagonal relaxation matrix,
\begin{equation}
{\bf{S}}^f=diag(s_0^f,s_1^f,s_2^f,s_3^f,s_4^f,s_5^f,s_6^f,s_7^f,s_8^f),
\end{equation}
where $0\leq s_i^f<2$. To simplify the LB algorithm, one part of the
diffusion term in Eq. (7) is treat as the source term here, and then
a novel discrete source term is introduced as
\begin{equation}
{F}_i=[\textbf{M}^{-1}(\textbf{I}-{\textbf{S}^f \over
2})\textbf{M}]_{ij}{R_j},
\end{equation}
where $\textbf{I}$ is the unit matrix, and $R_i$ is defined by
\begin{equation}
R_i={\omega_i c_{i\alpha}[\partial_t(r\phi u_\alpha+M\phi
\delta_{\alpha r})+c_s^2 rn_\alpha\lambda]\over{c_s^2}}.
\end{equation}
In the present model, the phase field variable $\phi$ is derived by
the summation of the distribution function $f_i$, and then is
computed by
\begin{equation}
\phi=\frac{1}{r}\sum\limits_i {{f_i}}.
\end{equation}
Physically, the density distribution in a multiphase system is
consistent with that of the phase field variable. To satisfy this
property, we take  the linear interpolation scheme to determine the
fluid density,
\begin{equation}
\rho=\phi\rho_l+(1-\phi)\rho_g,
\end{equation}
where $\rho_l$ and $\rho_g$ denote the liquid and gas fluid
densities, respectively. We also carried out the Chapman-Enskog
analysis to demonstrate the consistency of the present model with
the target equation (7). Applying the multiscale expansions to Eq.
(9), it is found in Appendix $\textbf{A}$ that the axisymmetric AC equation can
be recovered exactly from the present model, and the mobility is
given by
\begin{equation}
M= c_s^2 \delta_t(\tau_f-\frac{1}{2}),
\end{equation}
where $1/\tau_f={s_3^f}={s_5^f}$. In comparison with the standard LB
model~\cite{WangHL} for AC equation, it is found that the introduced
source terms to account for the axisymmetric effect include no
additional gradient in the present model.

We now give a discussion on implementation of the MRT model.
Generally, the MRT-LB equation (9) can be solved in two steps,
including the collision process,
\begin{equation}
f_i^{+}=f_i(\textbf{x},t)-({{{\bf{M}}^{-1}}{{\bf{S}}^f}{\bf{M}})_{ij}}[f_j(\textbf{x},t)-f_j^{eq}(\textbf{x},t)]+\delta_t{F}_i(\textbf{x},t)
\end{equation}
and the propagation process,
\begin{equation}
f_i(\textbf{x}+\textbf{c}_i\delta_t,t+\delta_t)=f_i^{+}.
\end{equation}
To reduce the matrix operations, it is wise that the collision
process of MRT model is conducted in the moment space. By
premultiplying the matrix $\textbf{M}$, the equilibrium distribution
function in moment space is derived as
\begin{equation}
\textbf{ Mf}^{eq}=\left(r\phi, -2r\phi, r\phi, \frac{r\phi{u_z}}{c},
-\frac{r\phi{u_z}}{c},\frac{r\phi{u_r}+M\phi}{c},-\frac{r\phi{u_r}+M\phi}{c},0,0\right)^T,
\end{equation}
where $u_r$ and $u_z$ are the radial and axial components of
velocity. Similarly, the discrete source term $R_i$ in the moment
space can also be obtained as
\begin{equation}
\textbf{ MR}=(0, 0, 0, mR_1,-mR_1, mR_2,-mR_2,0,0)^T,
\end{equation}
where $mR_1$ and $mR_2$ are given by
\begin{equation}
mR_1=\frac{\partial_t({r\phi{u_z}})+c_s^2r{n_z}\lambda}{c},
mR_2=\frac{\partial_t({r\phi{u_r}}+M\phi)+c_s^2r{n_r}\lambda}{c}.
\end{equation}
At the end of this subsection, we also would like to stress that the
present MRT model for the axisymmetric AC equation can reduce to the
SRT counterpart when the relaxation factor $s_i^f$ in Eq. (13)
equals to each other, and thus the SRT model is only one special
case of the MRT model. In this case, the more freedom in the choice
of the relaxation factors can provide more potential for the MRT
model to achieve better numerical accuracy and stability.

\subsection{ Axisymmetric LB model for the Navier-Stokes equations}

The LB evolution equation with a generalized collision matrix for
the axisymmetric NS equations can be written as
\begin{equation}
g_i(\textbf{x}+\textbf{c}_i\delta_t,t+\delta_t)-g_i(\textbf{x},t)=-\left[{{\bf{M}}^{
-1}}{{\bf{S}}^g}{\bf{M}}\right]_{ij}\left[g_j(\textbf{x},t)-g_j^{eq}(\textbf{x},t)\right]
+\delta_t{G}_i(\textbf{x},t),
\end{equation}
where $g_i$ is the hydrodynamic distribution function, $g_i^{eq}$ is
the corresponding equilibrium distribution function. To incorporate
the axisymmetric effect, a modified form of the equilibrium
distribution function is used~\cite{Liang2},
\begin{equation}
 {g_i^{eq}}=\left\{
\begin{array}{ll}
{rp \over {c_s^2}}({\omega _i} - 1) + r\rho{s_i}({\bf{u}}),  & \textrm{ $i=0$}    \\
{rp \over {c_s^2}}{\omega _i} + r\rho{s_i}({\bf{u}}),        & \textrm{ $i\neq0$} \\
\end{array}
\right.
\end{equation}
with
\begin{equation}
{s_i}({\bf{u}}) = {\omega _i}\left[ {{{{{\bf{c}}_i} \cdot {\bf{u}}}
\over {c_s^2}} + {{{{({{\bf{c}}_i} \cdot {\bf{u}})}^2}} \over
{2c_s^4}} - {{{\bf{u}} \cdot {\bf{u}}} \over {2c_s^2}}} \right].
\end{equation}
For fluids exposed to forces, the discrete lattice effects should be
taken into account, when the forces are introduced in LB
approach~\cite{Guo2}, and then the discrete force term in the MRT
framework for hydrodynamics is defined by
\begin{equation}
G_i=\left[\textbf{M}^{-1}(\textbf{I}-{\textbf{S}^g \over
2})\textbf{M}\right]_{ij}{T_j},
\end{equation}
where $T_i$ is the forcing distribution function, and ${\bf{S}}^g$
is a non-negative diagonal matrix given by
\begin{equation}
{\bf{S}}^g=diag(s_0^g,s_1^g,s_2^g,s_3^g,s_4^g,s_5^g,s_6^g,s_7^g,s_8^g).
\end{equation}
Different from other axisymmetric LB multiphase models~\cite{Liang2,
Premnath, Huang, Mukherjee, HuangJJ, Srivastava, LiuH}, a much
simplified forcing distribution function is constructed in this
model,
\begin{equation}
T_i=\omega_i\left[\frac{c_{i\alpha}{F}_\alpha}{c_s^2}+\frac{u_\alpha
\partial_\beta (r\rho)c_{i\alpha}c_{i\beta}}{c_s^2}-\rho u_r\right],
\end{equation}
where ${F}_\alpha$ is the total force and is given by
\begin{equation}
F_\alpha=r(\widetilde{{F}}_{s\alpha}+{G}_{\alpha})+(p-\frac{2\rho\nu}{r}u_r)\delta_{\alpha
r}.
\end{equation}
Taking the first-order moment of the distribution function, the
fluid velocity in this model can be evaluated as
\begin{equation}
r\rho u_\alpha=\sum_i{c_{i\alpha}g_i}+0.5\delta_t{F_\alpha},
\end{equation}
which can be further recast explicitly,
\begin{equation}
u_\alpha=\frac{ \sum_i{c_{i\alpha}g_i}+0.5\delta_t
[r(\tilde{{F}}_{s\alpha}+{G}_{\alpha})+ p\delta_{\alpha r}] }{
r\rho+\delta_t r^{-1}\nu\rho\delta_{\alpha r}}.
\end{equation}
The hydrodynamic pressure $p$ can be calculated in a particular
manner. As shown in Appendix $\textbf{C}$, it can be evaluated as
\begin{equation}
p=\frac{c_s^2}{1-\omega_0}\left[
\frac{1}{r}\sum_{i\neq0}g_i+\frac{\delta_t}{2}\textbf{u}\cdot\nabla\rho+\rho
s_0(\textbf{u})- \frac{{\delta_t}{s_p}\omega_0\rho u_r}{r}\right],
\end{equation}
where $s_p$ is a parameter relating to the relaxation times,
$s_p={{{s_1^g} + {s_2^g}-{s_1^g}{s_2^g}}\over{2{s_1^g}{s_2^g}}}$. We
also conducted the Chapman-Enskog analysis on the present model for
axisymmetric NS equations, and it is demonstrated in Appendix $\textbf{B}$ that
the correct hydrodynamic equations in the cylindrical coordinate
system can also be derived from the present model, and the relation
between the kinematic viscosity and the relaxation factor can be
presented as
\begin{equation}
\nu= c_s^2 \delta_t(\tau_g-\frac{1}{2} ),
\end{equation}
where $1/\tau_g=s_1^g=s_7^g=s_8^g$. Generally, the relation $s_7^g=s_8^g$ is satisfied in previous MRT model for hydrodynamics.
Here this new constraint for the relaxation factors is derived to account for the axisymmetric effect. In a multiphase system, the fluid
viscosity is no longer uniform due to its sharp jump at the
liquid-gas interface. For simplicity, the popular linear
interpolation scheme is used to determine the viscosity at the
interface~\cite{He},
\begin{equation}
\nu=\phi \nu_l+(1-\phi)\nu_g,
\end{equation}
where $\nu_l$ and $\nu_g$ represent the kinematic viscosities of the
liquid and gas phases, respectively. As for the MRT hydrodynamic
model, the collision process can also be implemented in moment
space. With some algebraic manipulations, the equilibrium and
forcing distribution functions in the moment space can be
respectively given by
\begin{equation}
\textbf{Mg}^{eq}=r\left(0,\frac{2p+\rho\textbf{u}^2}{c_s^2},-\frac{3p+\rho\textbf{u}^2}{c_s^2},\frac{\rho{u_z}}{c},-\frac{\rho{u_z}}{c},\frac{\rho{u_r}}{c},-\frac{\rho{u_r}}{c},
\frac{\rho(u_z^2-u_r^2)}{c^2},\frac{\rho{u_r}{u_z}}{c^2}\right)^T
\end{equation}
and $\textbf{MT}=$
\begin{equation}
\left(r{u_\alpha}\partial_\alpha\rho,2\rho{u_r},-u_\alpha\partial_\alpha{r\rho}-\rho{u_r},\frac{F_z}{c},-\frac{F_z}{c},
\frac{F_r}{c},-\frac{F_r}{c},\frac{2}{3}({u_z\partial_z{r\rho}-u_r\partial_r{r\rho}}),
\frac{1}{3}({u_z\partial_r{r\rho}+u_r\partial_z{r\rho}})\right)^T.
\end{equation}

In practice, the time derivative and the spatial gradients should be
discretized with suitable difference schemes. In this paper, we
adopt the explicit Euler scheme to calculate the time derivative as
follows,
\begin{equation}
\partial_t{\chi(\textbf{x},t)}=\frac{\chi(\textbf{x},t)-\chi(\textbf{x},t-\delta_t)}{\delta_t}.
\end{equation}
Following the work of Liang {\it{et al.}}~\cite{Liang2}, the
gradient and the Laplacian operator are discretized with the
following four-order isotropic difference schemes,
\begin{equation}
\nabla{\chi(\textbf{x},t)}=\sum_{i\neq0}\frac{ \omega_i
\textbf{c}_i[ 8\chi( \textbf{x}+\textbf{c}_i \delta_t,t )-\chi(
\textbf{x}+2\textbf{c}_i \delta_t,t ) ] }{6c_s^2 \delta_t },
\end{equation}
and
\begin{equation}
\nabla^2{\chi(\textbf{x},t)}=\sum_{i\neq0}\frac{ \omega_i[ 16\chi(
\textbf{x}+\textbf{c}_i \delta_t,t )-\chi( \textbf{x}+2\textbf{c}_i
\delta_t,t )-15\chi(\textbf{x},t) ] }{6c_s^2 \delta_t^2 },
\end{equation}
where $\chi$ represents an arbitrary variable. The evaluation of the
gradient terms at fluid nodes neighbouring to boundary, using Eqs.
(39) and (40), typically requires unknown information at the ghost
nodes, and we use the mirror symmetric rule to derive the unknown
values for fluid nodes neighbouring to solid wall, while applying
the axis-symmetric means for the axial nodes. In addition, the
boundary conditions for the distribution functions should also be
specially treat, since the singularity arises at the symmetry axis
of $r=0$. To avoid this problem, we set the first lattice line at
$r=0.5\delta_x$, and apply the symmetry boundary condition for axial
boundary. For other solid boundary, we impose the no-slip bounce
back boundary condition. The detailed treatments on the used
boundary conditions and discretization schemes can be also referred to
Refs.~\cite{Guo1, Liang2}.

\section{NUMERICAL VALIDATION FOR AXISYMMETRIC LB MULTIPHASE  MODEL}\label{sec:result}
In this section, we will test the accuracy and stability of the
proposed axisymmetric LB model by using several basic tests. These
typical numerical examples include the static droplet,
oscillation of a droplet, breakup of a liquid thread, and bubble
rising in a continuous phase, where a very board range of density
ratios are considered to highlight the advantage of the present LB
model.
\subsection{Static droplet test}
A benchmark problem of the static droplet is first used to verify
the present LB model for axisymmetric multiphase flows. Initially, a
stationary droplet is located at the computational domain with the
size of $N_z\times N_r=200\times100$, which is centered at the node
$(100,0)$ and its radius ($R$) occupies 50 lattice units. The left
and right boundary conditions for both $f_i$ and $g_i$ are set to be
periodic, while the symmetry boundary condition is applied at the
axial line and the no-slip boundary condition is imposed at the
upper boundary. To be smooth across the interface, the phase field
variable is initialized by
\begin{equation}
\phi(z,r)=0.5+0.5\tanh\left[2\frac{R-\sqrt{(z-100)^2+r^2}}{W}\right],
\end{equation}
and the corresponding distribution of the density field based on Eq.
(17) can be given by
\begin{equation}
\rho(z,r)=\frac{\rho_l+\rho_g}{2}+\frac{\rho_l-\rho_g}{2}\tanh\left[2\frac{R-\sqrt{(z-100)^2+r^2}}{W}\right],
\end{equation}
where $\rho_l$ and $\rho_g$ in the test are set to be 1000 and 1,
corresponding to a large density ratio of 1000. Some other
simulation parameters are given as follows: $W=4.0$, $\sigma=0.001$,
$M=0.01$ and $\nu_l=\nu_g=0.1$. Considering the components of the
equilibrium in moment space, we fix $s_0^f=s_7^f=s_8^f=1$ as usual,
and the relations $s_1^f=s_2^f$ and $s_3^f=s_4^f=s_5^f=s_6^f$ are
satisfied. For simplicity, the values of $s_1^f$ and $s_2^f$ in the
simulation are set to be 1. The relaxation factors $s_3^f$ and
$s_5^f$ are given by the value of $1/\tau_f$, which is further
determined by the mobility $M$ based on Eq. (18). As for the ones in
the matrix $\textbf{S}^g$, the relaxation times $s_i^g$ are chosen
to be unity expect for $s_1^g$, $s_7^g$ and $s_8^g$, which are adjusted
according to the fluid viscosity.

\begin{figure}
\centering
\includegraphics[width=2.65in,height=2.5in] {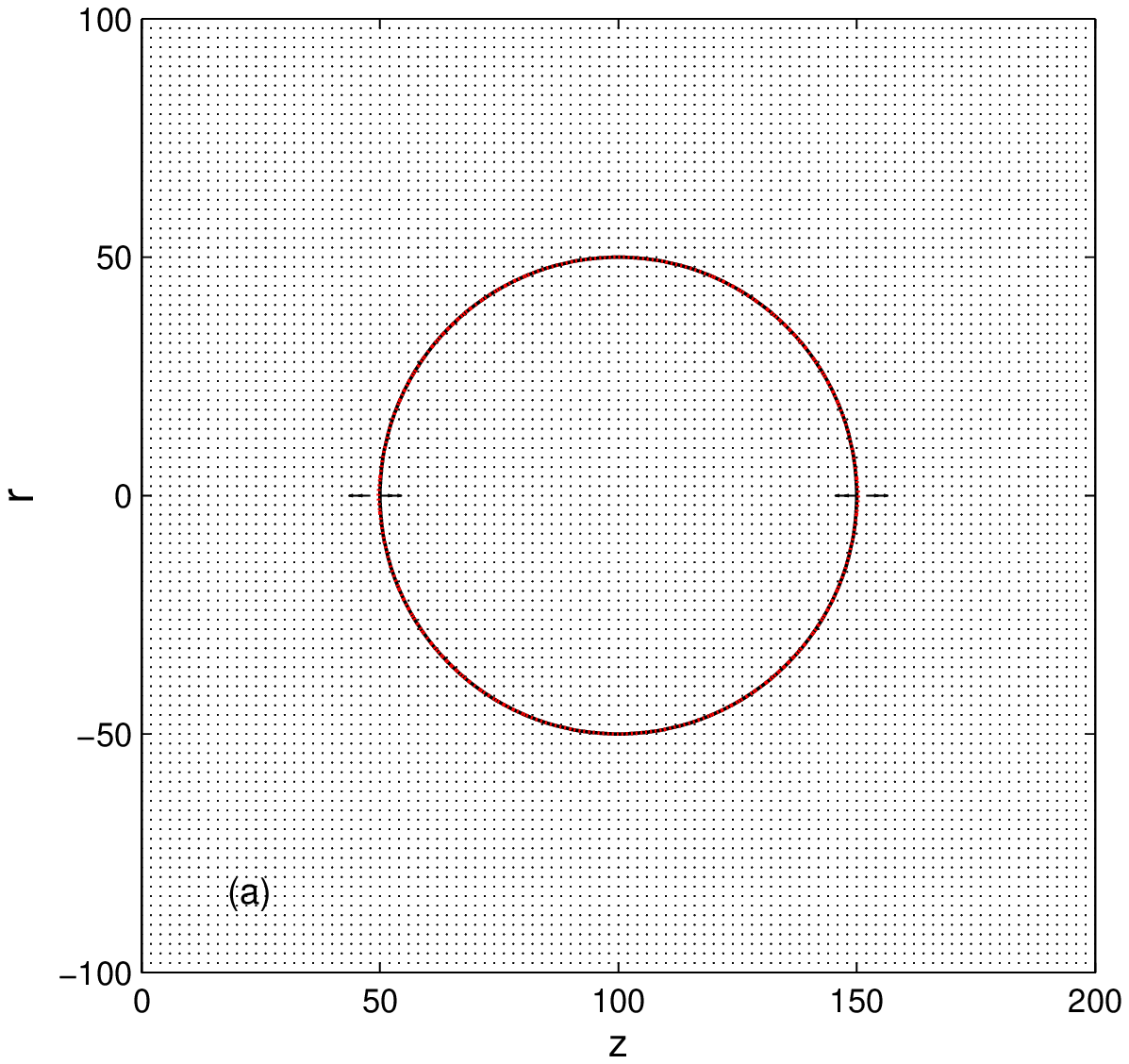}~~~
\includegraphics[width=2.65in,height=2.5in] {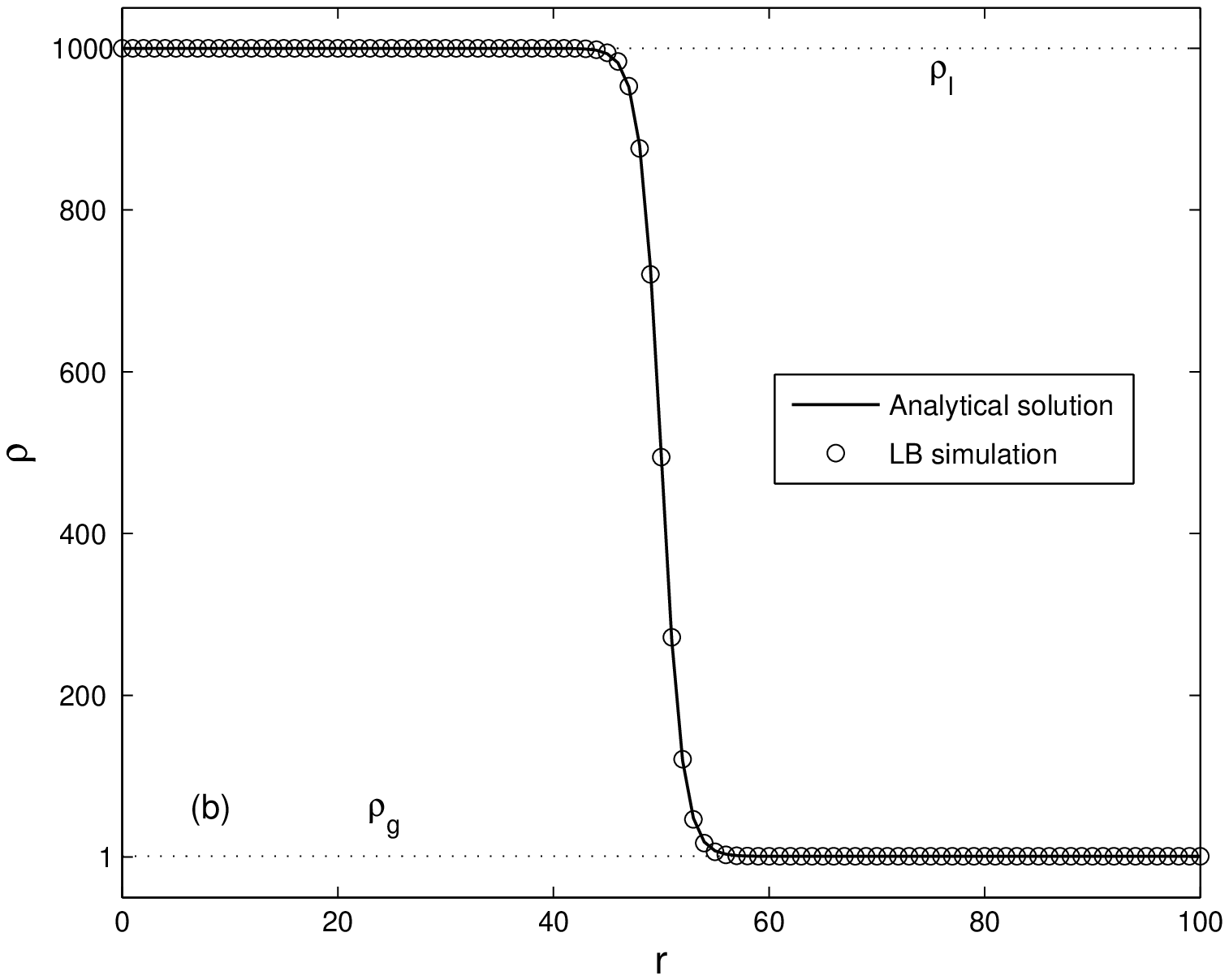}\\
\tiny\caption{ static droplet test with density ratio
$\rho_l/\rho_g=1000$: (a) the velocity distribution of the whole domain at the equilibrium state and the solid and dashed lines respectively represent the equilibrium shape of the droplet and its initial shape; (b) density profile across the interface obtained from LB simulation and corresponding analytical solution.}
\end{figure}

Figure 1(a) depicts the steady droplet shape obtained by the present
LB model, together with the initial one. It is found that they match
to each other with a high accuracy. Further, we also plotted the
numerical prediction of the density profile across the interface in
Fig. 1(b), where the analytical solution given by Eq. (42) is as
well presented  for the comparison. It can be clearly observed that
numerical result agrees well with the analytical solution, which
indicates that the present model can correctly solve the phase
interface. The existence of the spurious velocities is a common
undesirable feature in many numerical methods for multiphase flows.
If they have comparable magnitudes as the real fluid velocity,
numerical instability or some unphysical phenomena could take place,
and thus achieving small spurious velocities is very significative.
The spurious velocities also rooted in the LB method arise from the
imbalance between discrete forces in the interfacial
region and cannot be totally eliminated to round-off in
the framework of the LB approach~\cite{Guo3}. We also displayed the spurious
velocities generated by the present LB model. The distribution of
the velocity field in the whole computational domain is shown in
Fig. 1(a). It can be obviously found that the spurious velocities
indeed exist in the vicinity of the interface, and the maximum
magnitude of the spurious velocities computed by
$|\textbf{u}|_{max}=(\sqrt{u^2+v^2})_{max}$ is about
$1\times10^{-6}$. We also compared the spurious velocities generated by different LB approaches.
The axisymmetric color-gradient LB multiphase model proposed by Liu {\it{et al.}}~\cite{LiuH} can
only deal with multiphase flows with moderate density ratio and they reported that it can derive the spurious
velocities at the level of $10^{-5}$. In addition, it is found that the maximum amplitude of spurious velocities
in an improved pseudo-potential model~\cite{Yu} has the order of $10^{-3}$. In comparisons with these common LB models, we
can conclude that the present phase-field based LB model can derive relatively low
spurious velocities.

\subsection{Droplet oscillation test}

The droplet oscillation is a classic example that is widely used to
validate axisymmetric multiphase LB model for simulating dynamic
problem~\cite{Premnath, Srivastava, Liang2, LiuH}. The liquid
droplet will exhibit oscillatory behavior, if it is initially
distorted from equilibrium spherical shape into an elliptical one.
We intend to compare the droplet oscillation frequency obtained from
the LB simulation with the analytical solution reported by Miller
and Scriven~\cite{Miller}. According to their analysis, the
theoretical prediction for the oscillation frequency of the $n$th
mode can be given by
\begin{equation}
\omega_n=\omega_n^*-0.5\alpha\sqrt{\omega_n^*}+0.25\alpha^2,
\end{equation}
where $\omega_n^*$ is Lamb's natural resonance frequency,
\begin{equation}
\omega_n^*=\sqrt{ \frac{n(n-1)(n+1)(n+2)}{
R_e^3[n\rho_g+(n+1)\rho_l] }\sigma },
\end{equation}
where $R_e$ is the radius of droplet at the equilibrium state. In
Eq. (43), the parameter $\alpha$ is included to account for the
viscosity contribution and is expressed by~\cite{Miller}
\begin{equation}
\alpha=\frac{ (2n+1)^{2}\rho_l\rho_g\sqrt{\nu_l\nu_g} }{
\sqrt{2}R_e[n\rho_g+(n+1)\rho_l](\rho_l\sqrt{\nu_l}+\rho_g\sqrt{\nu_g})}.
\end{equation}
The mode of the oscillation is denoted by $n$, which is set to be 2
for an initial ellipsoidal shape considered here.

The initial computational setup consists of an ellipsoidal droplet
with the half -axial and -radial lengths denoted by $R_z$ and $R_r$,
placed in a domain with the size of $N_z\times N_r=300\times150$.
The boundary conditions are chosen as those in the static droplet
test. To match this initial condition, the profile of the phase
field variable is given by
\begin{equation}
\phi(z,r)=0.5+0.5\tanh\left[2R_e\frac{1-\sqrt{(z-z_c)^2/R_z^2+(r-r_c)^2/R_r^2}}{W}\right],
\end{equation}
where $(z_c,r_c)=(150,0)$ is the center coordinate of ellipsoidal
droplet, and the equilibrium droplet radius $R_e$ can be calculated
as $(R_r^{2}R_z)^{1/3}$ based on the mass conservation of the liquid
droplet. The density ratios used in this test range from 10 and 100,
and we find that the model is still stable for a very large density
ratio of 1000. Considering a very low oscillation frequency at a
large density ratio, it is generally measured difficultly, and thus
the result in this case is not presented. Some remaining physical
parameters in the simulation are fixed as $W=4$, $\sigma=0.3$,
$M=0.01$, $\nu_l=\nu_g=0.1$, and the relaxation factors in the
matrices $\textbf{S}^f$ and $\textbf{S}^g$ are set to be those in
the last test.

\begin{figure}
\centering
\includegraphics[width=1.93in,height=1.9in]{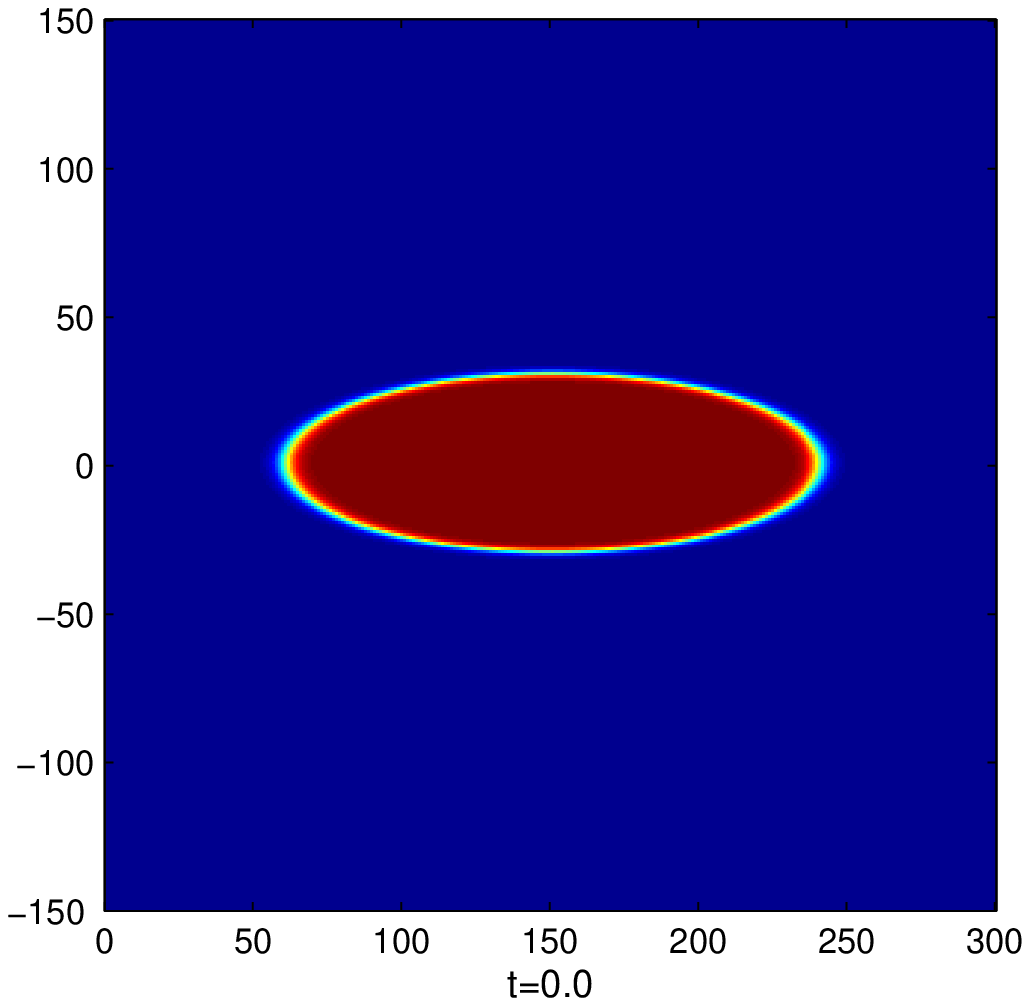}~~~~
\includegraphics[width=1.93in,height=1.9in]{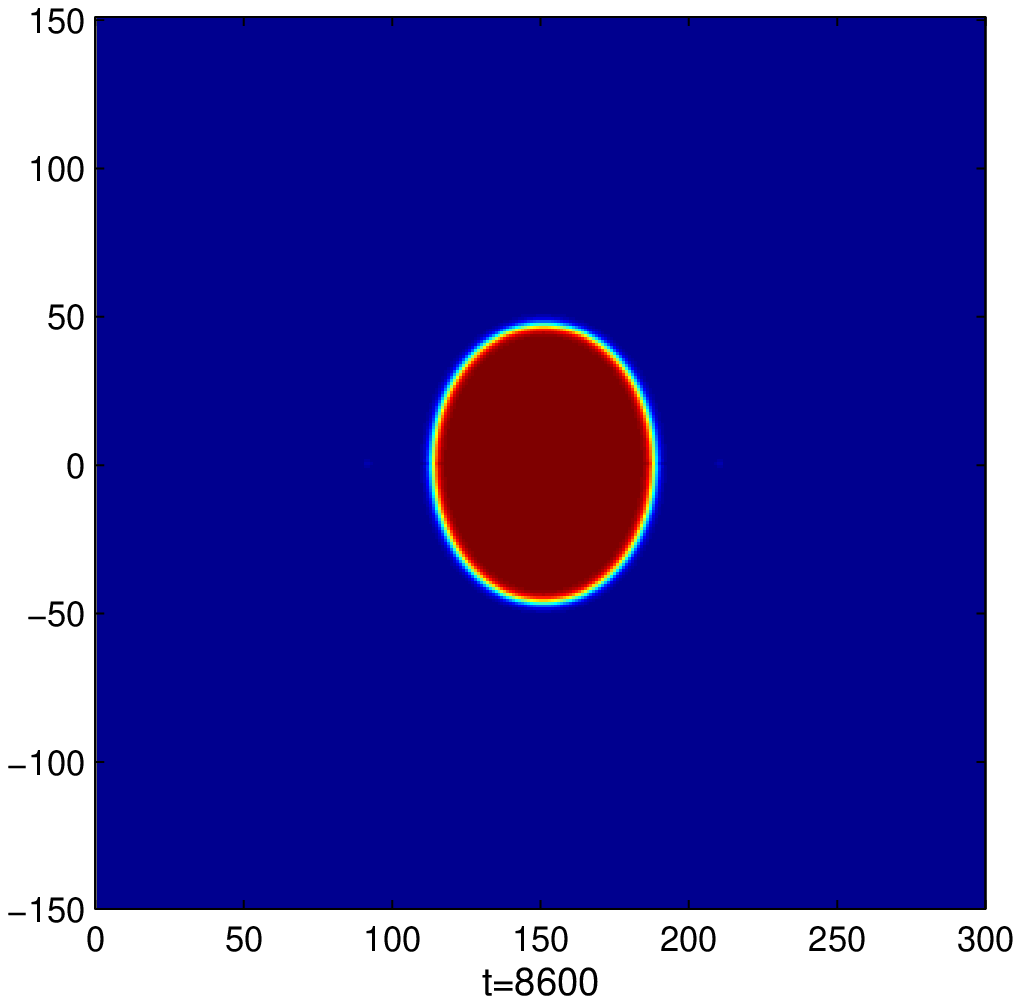}\\
\includegraphics[width=1.93in,height=1.9in]{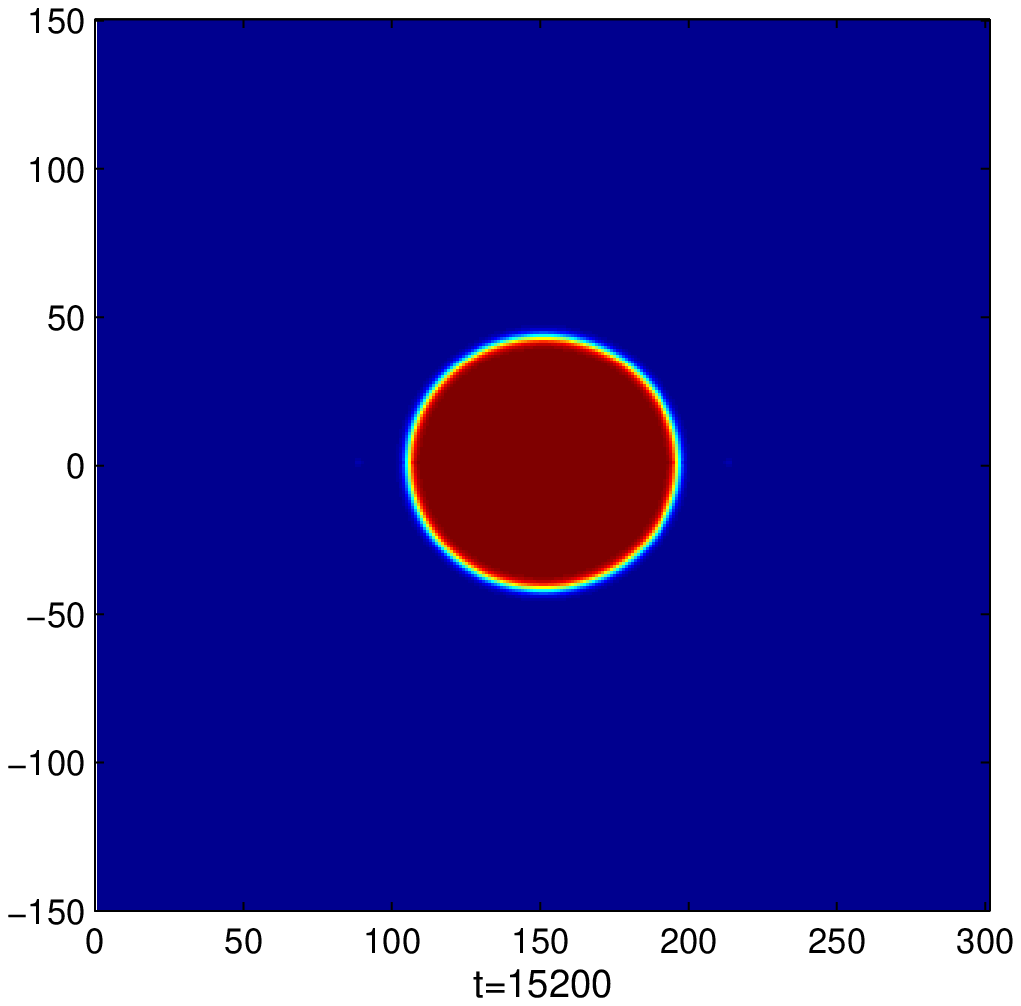}~~~~
\includegraphics[width=1.93in,height=1.9in]{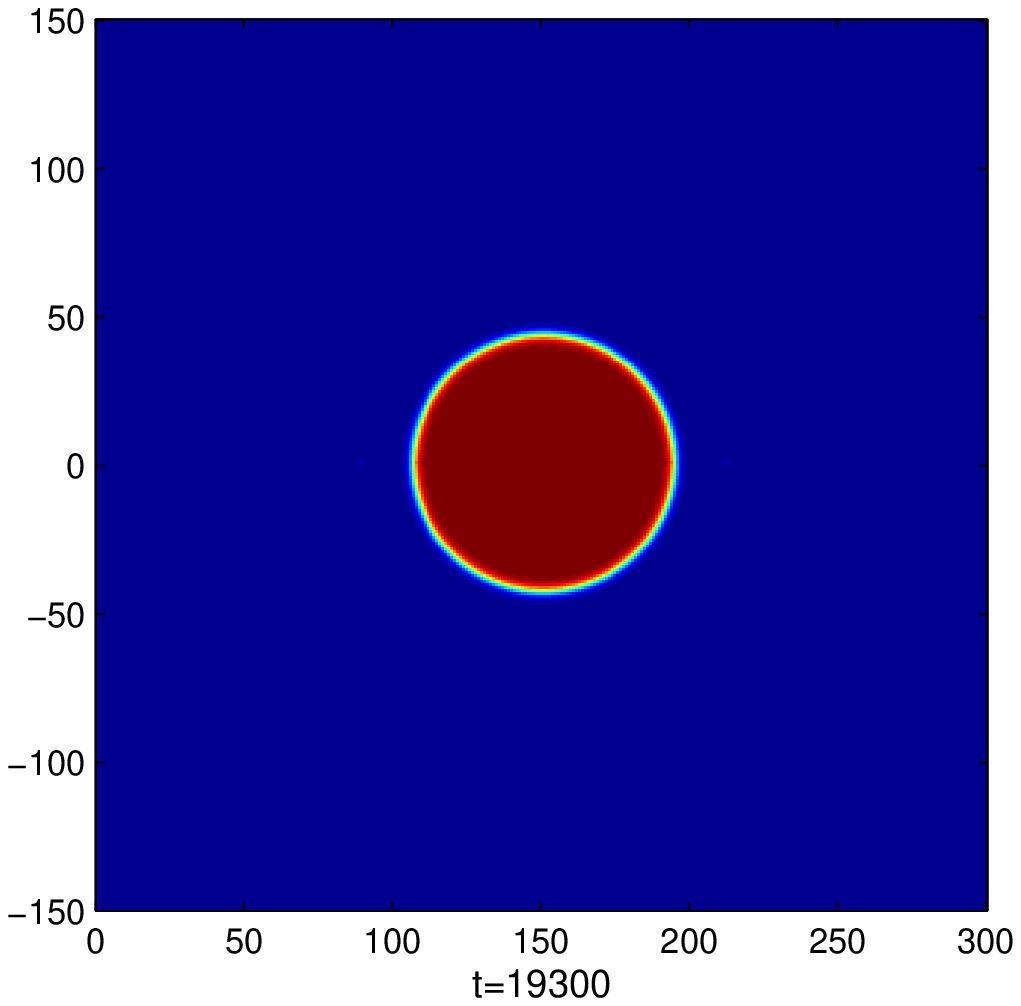}
 \tiny\caption{Evolution of the shape of an oscillating droplet with $\rho_l/\rho_g=100$, $R_r=30$, $R_z=90$, $\sigma=0.3$. }
\end{figure}

\begin{figure}
\centering
\includegraphics[width=3.6in,height=2.8in]{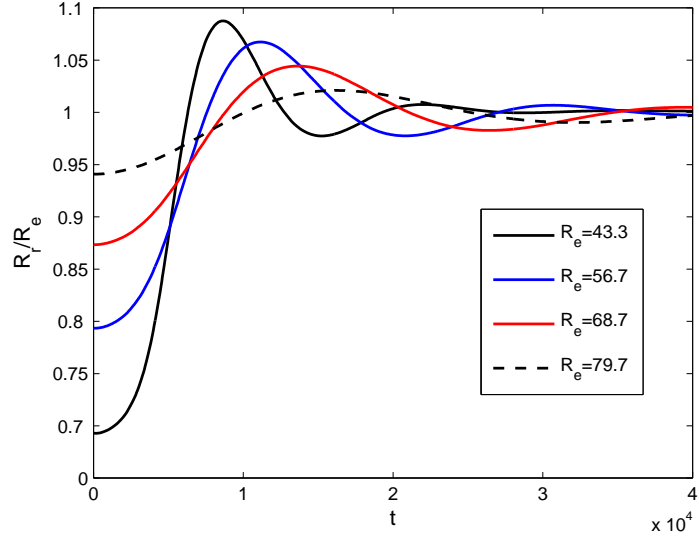}
 \tiny\caption{Time evolution of the half-axis length $R_r$ at different equilibrium radii with $\rho_l/\rho_g=100$, $\sigma=0.3$. And the half-axis length $R_r$ is normalized by the equilibrium radius $R_e$. }
\end{figure}
\begin{figure}
\centering
\includegraphics[width=3.2in,height=2.55in]{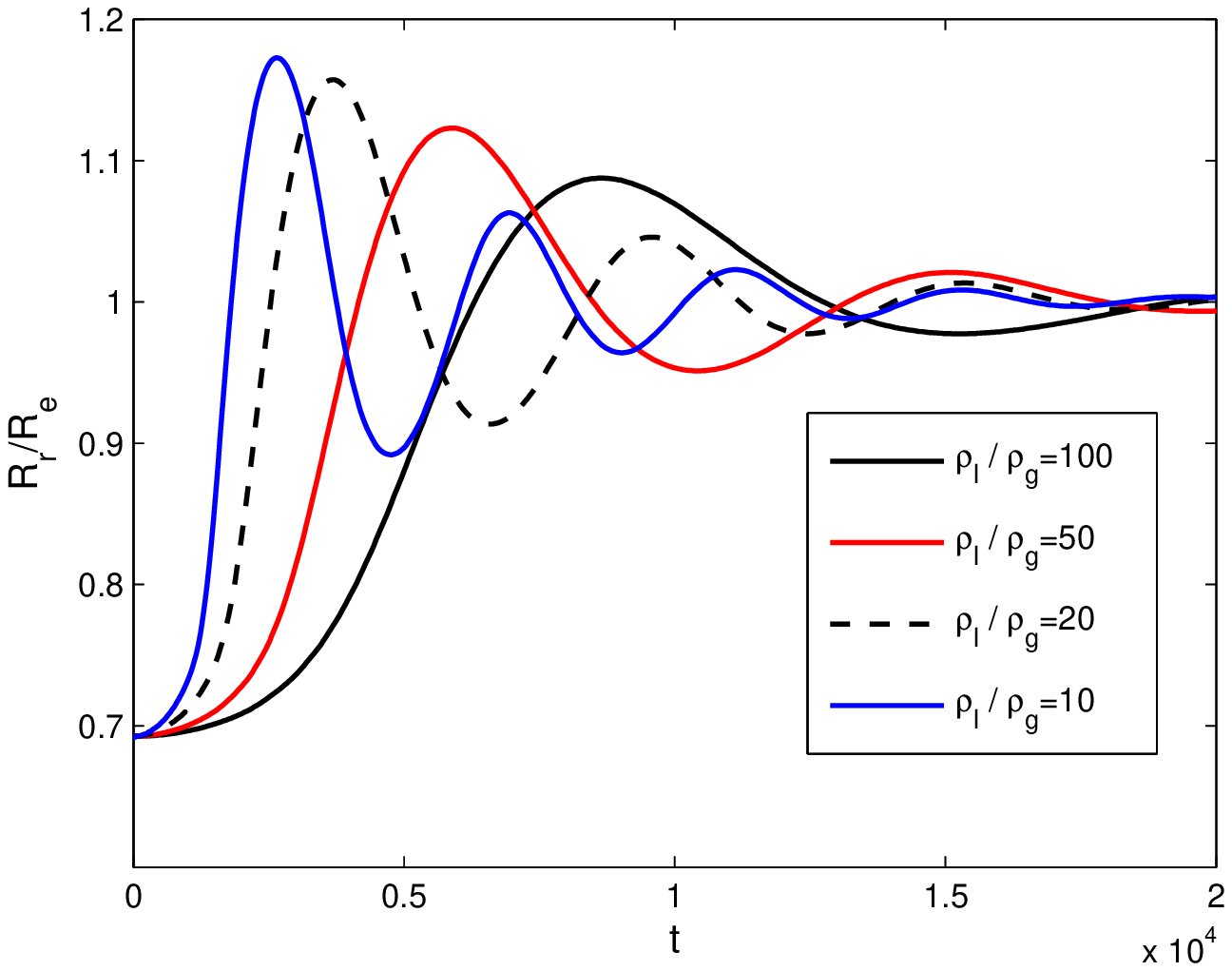}
 \tiny\caption{Time evolution of the half-axis length $R_r$ at different density ratios with $R_r=30$, $R_z=90$, $\sigma=0.3$. And the half-axis length $R_r$ is normalized by the equilibrium radius $R_e$.}
\end{figure}

Figure 2 shows the snapshots of an oscillating droplet at different
times, where the density ratio is set as 100 and the initial droplet
size is given by $R_z=90$ and $R_r=30$. As can be seen from Fig. 2,
the droplet changes from a prolate shape at initial time to the
oblate one at $t=8600$. Such change continues until the droplet
reaches its equilibrium spherical shape. The above droplet behaviors
are consistent with the expectation. We also conducted a
quantitative study of droplet oscillation problem. The variation of
the half-axis length $R_r$ versus time is plotted in Fig. 3, where
the results with other initially given values of $R_r=45$, 60 and 75
are also presented to examine the effect of droplet size. It can be
found from Fig. 3 that the amplitude of the oscillation normalized
by the corresponding equilibrium radius $R_e$ fluctuates around 1
for all cases, and as expected, the dimensionless maximum amplitude
is reduced for a larger droplet size. Further, from Fig. 3 we can
compute the droplet oscillating frequency and showed the results in
Table~\ref{Table 1}. For a comparison, the theoretical solution for
predicting the oscillating frequency was also presented. It is found
that they agree well in general, with the maximum relative error of
around 8.9\%.

\begin{table}
\centering  
\caption{Comparison between the computed oscillating frequency $\omega_{LB}$ and corresponding analytical values $\omega_{ana}$ at different equilibrium radii for $\rho_l/\rho_g=100$, $\sigma=0.3$ and $R_z=90$.}\label{Table 1}
\setlength{\tabcolsep}{5.0mm}{ 
\begin{tabular}{lllll}
  \Xhline{0.8pt}
  $R_e$&43.3&56.7&68.7&79.7\\
  \Xhline{0.8pt}
  $\omega_{LB}(\times10^{-4})$ & 4.8934 & 3.4074 & 2.5964 & 2.1313\\
  $\omega_{ana}(\times10^{-4})$ & 5.3764 & 3.5864 & 2.6908 & 2.1533\\
  $E_r=\frac{|\omega_{LB}-\omega_{ana}|}{\omega_{ana}}\times100\% $ & 8.9\% & 5.0\% &3.5\% &5.7\%\\
  \Xhline{0.8pt}
\end{tabular} }
\end{table}

The influence of the density ratio on the oscillating frequency is
also investigated here. We use the present model to simulate this
case with four different density ratios $\rho_l/\rho_g=10$, 20, 50
and 100, and the droplet size is fixed to be $R_z=90$ and $R_r=30$.
Figure 4 depicts the evolution of the oscillating amplitude with
these typical density ratios. It can be clearly observed that the
derived dimensionless amplitudes also fluctuates around 1, but the
range is evidently reduced with the increase of the density ratio.
In addition, the numerical predictions of the oscillating frequency
obtained from the present LB model with various density ratios are
summarized in Table~\ref{Table2}, together with the corresponding analytical
results. It can be seen that increasing density ratio decreases the
droplet oscillating frequency, and the computed oscillating
frequencies show good agreement with the analytical results for all
density ratios, with a maximum error of about 8.9\%.
\begin{table}
\centering  
\caption{Comparison between the computed oscillating frequency $\omega_{LB}$ and corresponding analytical values $\omega_{ana}$ at different density ratios for $R_z=90$, $R_r=30$, and $\sigma=0.3$.}\label{Table2}
\setlength{\tabcolsep}{4.5mm}{ 
\begin{tabular}{lllll}
  \Xhline{0.8pt}
  $\rho_l/\rho_g$ & 10 & 20 & 50 & 100\\
  \Xhline{0.8pt}
  $\omega_{LB}(\times10^{-4})$ & 15.2504 & 11.0000 & 7.0439 & 4.8934\\
  $\omega_{ana}(\times10^{-4})$ & 16.5854 & 12.0000 & 7.5895 & 5.3764\\
  $E_r=\frac{|\omega_{LB}-\omega_{ana}|}{\omega_{ana}}\times100\% $ & 8.0\% & 8.3\% & 7.2\% & 8.9\%\\
  \Xhline{0.8pt}
\end{tabular} }
\end{table}

\subsection{Breakup of a liquid thread test}
To show the capacity of the current model in simulating large
interface topological changes, in the subsection we consider a
fascinating problem of the breakup of a liquid thread into satellite
droplets. The breakup of liquid filaments is of long-standing
importance not only in its own interests, but also in the fact that
numerous practical applications, such as gene chip arraying, ink-jet
printing and microfluidics, depend critically on the knowledge of
the breakup mechanisms~\cite{Eggers, Wijshoff}. The earliest
experimental study on the breakup of liquid filament was attributed
to Plateau~\cite{Plateau}, and later Rayleigh~\cite{Rayleigh}
performed the linear stability analysis on this problem, who argued
that a cylindrical liquid thread of radius $R$ is unstable, if the
wavelength $\tilde{\lambda}$ of a disturbance on thread surface is
longer than its circumference $2\pi R$. Usually, this instability
criterion can also be described by using the wave number $k$, i.e.,
$k=2{\pi}R/\tilde{\lambda}$. When $k$ is smaller than 1, a liquid
thread is unstable and can separate into small droplets, while it is
stable for the case of $k>1$. Some preliminary numerical experiments
with $k>1$ were first performed, and the breakup phenomena indeed
cannot be observed in the simulations, which is in accordance with
the linear theory. In the following, we restrict our attention to
the breakup case, i.e., $k<1$, and compare the present numerical
results with some available data.

The simulations were carried out in a $N_z\times
N_r=\tilde{\lambda}\times 200$ lattice domain with the same boundary
conditions used in the static and oscillating droplet tests. An
initial disturbance imposed on thread surface is given by setting
the phase field profile as
\begin{equation}
\phi(z,r)=0.5+0.5\tanh{\frac{2(R+d-r)}{W}},
\end{equation}
where the radius $R$ is taken as 60, and $d$ is the perturbation
function given by $d=0.1R\cos(2\pi z/\tilde{\lambda})$. To examine
the wave number effect, several cases with different wavelengths
$\tilde{\lambda}$=420, 500, 600, 800, 1000, 1200, 1800 were
simulated, which correspond to various wave numbers $k$=0.90, 0.75,
0.63, 0.47, 0.38, 0.31, 0.21. In the simulations, two different
density ratios of $\rho_l/\rho_g=100$ and 10 are considered and the
interfacial tension $\sigma$ is fixed as 0.3. We would like to
stress that the model can also tackle this case with a high density
ratio of 1000, while it takes plenty of time for the system to reach
the breakup state under the identical surface tension force. In this
case, the moderate density ratio is merely considered. Here some other
used physical parameters are given as $W=4.0$, $M=0.01$ and
$\nu_l=\nu_g=0.1$. The parameters in both $\textbf{S}^f$ and
$\textbf{S}^g$ are chosen as those in the last test.
\begin{figure}
\centering
\includegraphics[width=1.0in, height=3.5in]{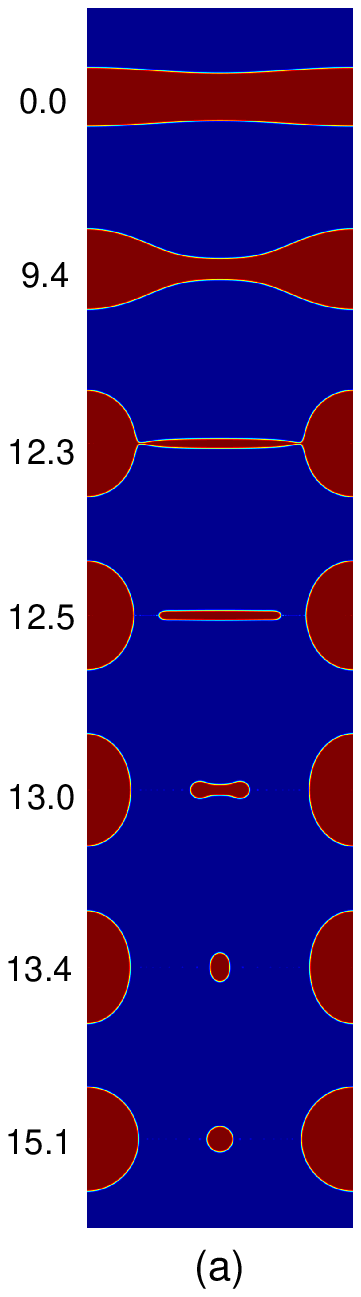}
\includegraphics[width=1.8in, height=3.5in]{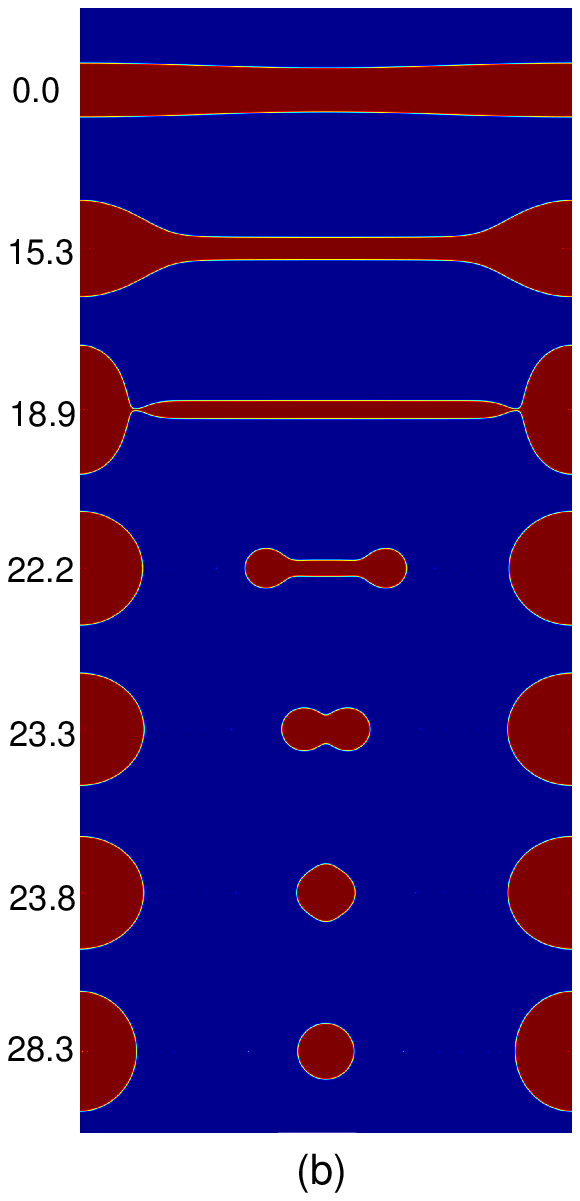}
\includegraphics[width=2.5in, height=3.5in]{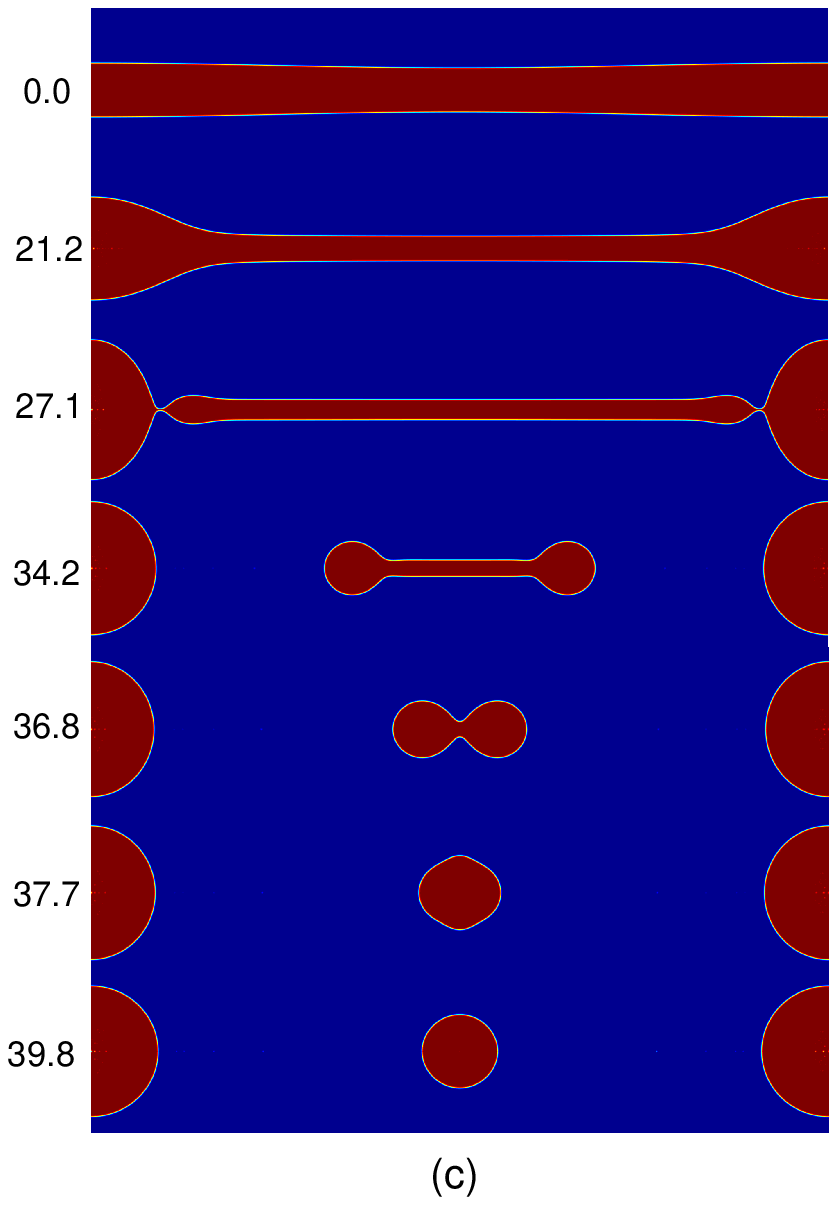}
 \tiny\caption{Evolutions of the breakup of a liquid thread at different wave numbers with the density ratio of $\rho_l/\rho_g=100$,
(a) $k=0.63$, (b) $k=0.31$, (c) $k=0.21$. Time has been normalized by the capillary time $\sqrt{R^3 \rho_l/\sigma}$. }
\end{figure}
We first present the results for the density ratio of 100. Figure 5
depicts the time evolution of a liquid thread with three typical
wave numbers, where time has been normalized by the capillary time
$\sqrt{R^3\rho_l/\sigma}$. It can be observed from Fig. 5 that for
all cases, the interfacial perturbation grows continuously at early
stage, and then the liquid thread in middle region becomes more and
more thin, while its ends are enlarged with time. As time goes on,
the liquid thread breaks up at two thin linking points, forming into
a liquid ligament as well as a main droplet. Afterwards, the liquid
ligament shrinks constantly with the action of the dominant surface
tension, until it reaches an equilibrium spherical state. Lastly, a
pair of liquid droplets including the main droplet and the satellite
one can be observed in the system. Actually, the formed liquid
ligament can be treat as the secondary liquid thread, and whether
its breakup occurs or not depends on the new wavelength. If the
wavelength is sufficiently large such that the instability criterion
can be satisfied, the secondary breakup of the ligament can take
place, leading to the formation of multiple satellite droplets. The
above behaviors of the thread breakup obtained by the present model
are qualitatively consistent with the previous
studies~\cite{Ashgriz, Tjahjadi}.
\begin{figure}
\centering
\includegraphics[width=1.0in, height=3.5in]{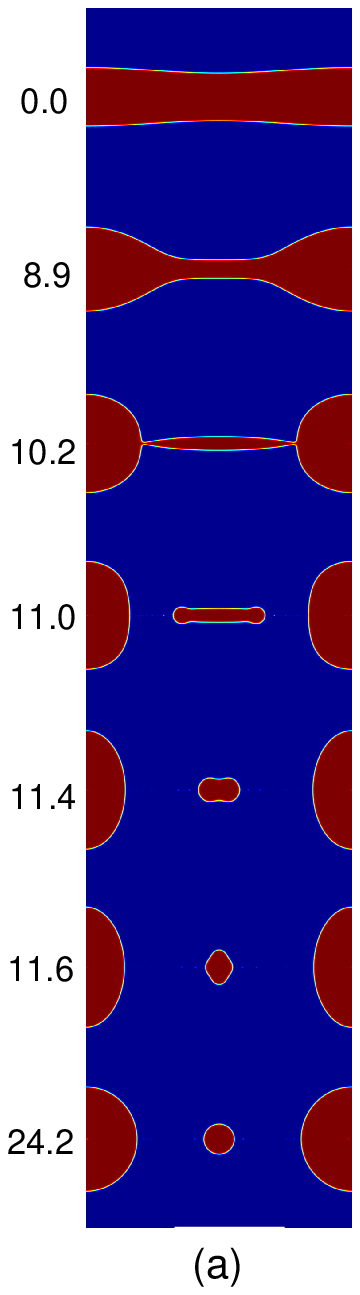}
\includegraphics[width=1.8in, height=3.5in]{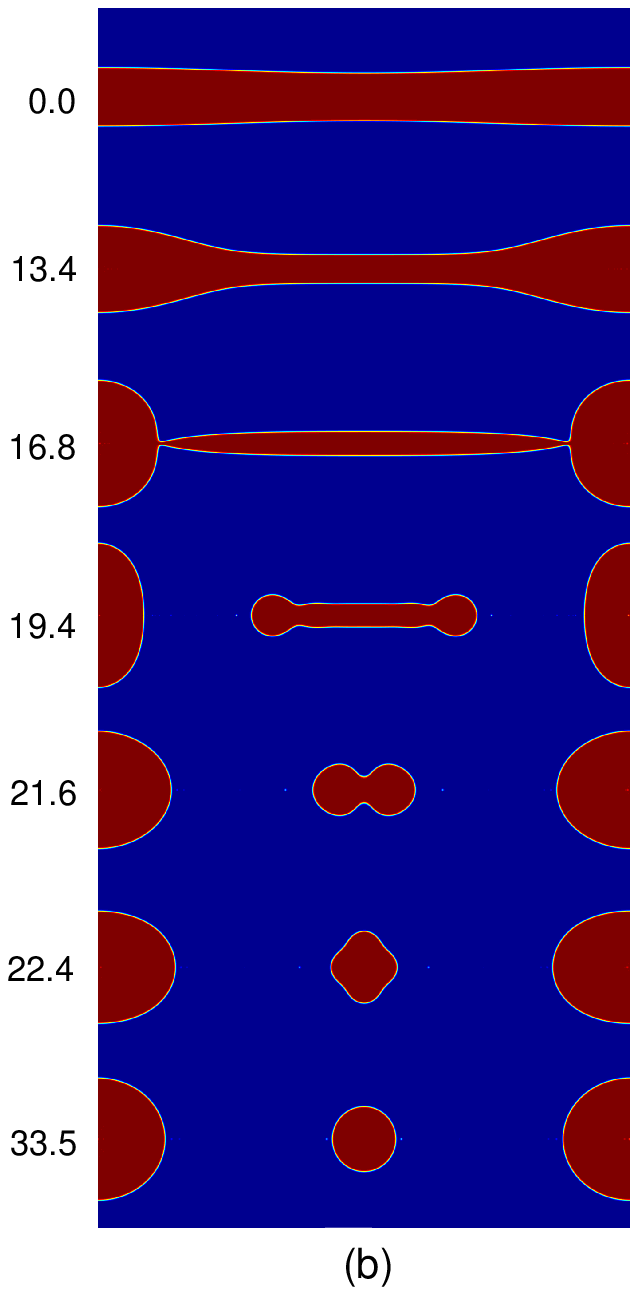}
\includegraphics[width=2.5in, height=3.5in]{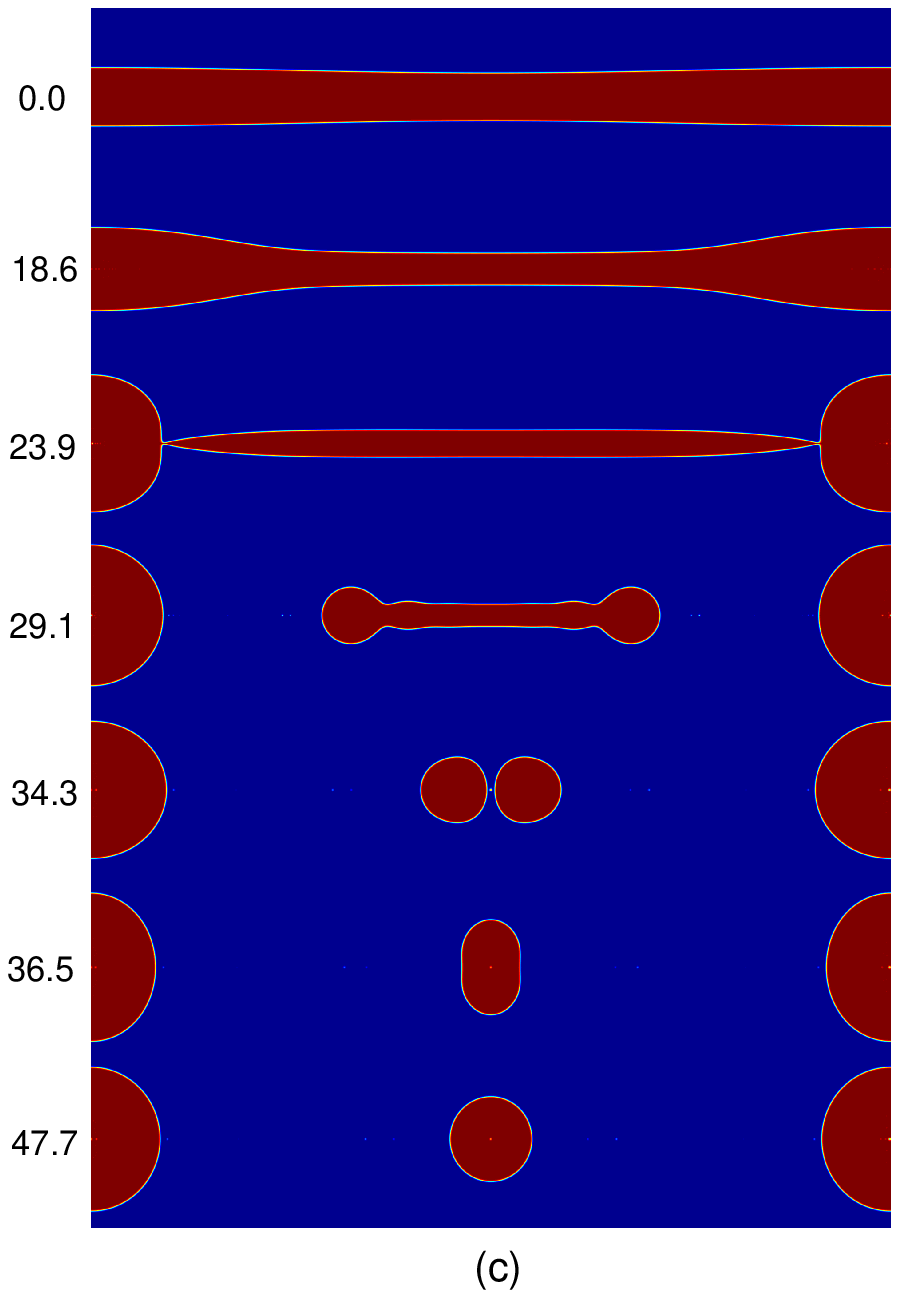}
 \tiny\caption{Evolutions of the breakup of a liquid thread at different wave numbers with the density ratio of $\rho_l/\rho_g=10$,
(a) $k=0.63$, (b) $k=0.31$, (c) $k=0.21$. Time has been normalized by the capillary time $\sqrt{R^3 \rho_l/\sigma}$. }
\end{figure}

To further investigate the density ratio effect, we also simulate
the breakup of a liquid thread with the density ratio of
$\rho_l/\rho_g=10$, where other physical parameters remain
unchanged. The snapshots of the breakup of a liquid thread with
three different wave numbers are shown in Fig. 6. It is found that
compared with the case of $\rho_l/\rho_g=100$, the liquid thread
exhibits similar behaviors at early time for all the wave numbers.
The initial disturbance imposed on the thread surface increases in
time, which then results in the formation of a thin ligament and a
main droplet. Afterwards, some distinct behaviors of a liquid
ligament are observed for different wave numbers. For the case of
$k=0.62$ or 0.31, the filament contracts continuously into one
satellite droplet. This phenomenon is a familiar spectacle in
simulation results with the density ratio of 100. Whereas for the
case of $k=0.21$, the liquid ligament shrinks at first, and then
pinches off due to the Rayleigh instability, which leads to the
relaxation of two daughter droplets. The two daughter droplets move
toward each other, and eventually merge into a larger satellite
droplet. These processes are the results of the inertia and surface
tension action, and are in line with the results reported in
literatures~\cite{Liang2, Ashgriz, Tjahjadi}. In contrast, the
secondary breakup of the ligament does not occur for the density
ratio of 100, which indicates that a higher density ratio prevents
the interface rupture. Besides, the breakup time is a concerned
physical quantity in the study of thread instability. We also
measured the breakup time for two density ratios with various wave
numbers. The computed breakup times for the density ratio of 100 at
three wave numbers $k=0.63$, 0.31, 0.21 are 12.3, 18.9, 27.1, while
they for the density ratio of 10 are 10.2, 16.8, 23.9. Therefore it is
concluded that increasing density ratio can slow down the breakup of a liquid thread.
Furthermore, we also give a quantitative study on the thread
breakup, and showed in Fig. 7 the main and satellite droplet sizes
with two density ratios and various wave numbers. For comparisons,
some available literature results, including the finite element
results of Ashgriz and Mashayek~\cite{Ashgriz}, the analytical
solutions and experimental data of Lafrance~\cite{Lafrance}, as well
as the experimental data by Rutland and Jameson~\cite{Ruland}, are
also presented. It can be seen that both the main and satellite
droplet sizes decrease with the increasing wave number, and also, it
is found that the satellite droplet size is reduced for a larger
density ratio, whereas the trend is just the reverse for the main
droplet due to the mass conservation. The numerical predictions for
the droplet sizes obtained from the present LB simulations are
further found to be in good agreement with these available date in
general.
\begin{figure}
\centering
\includegraphics[width=3.6in,height=2.8in]{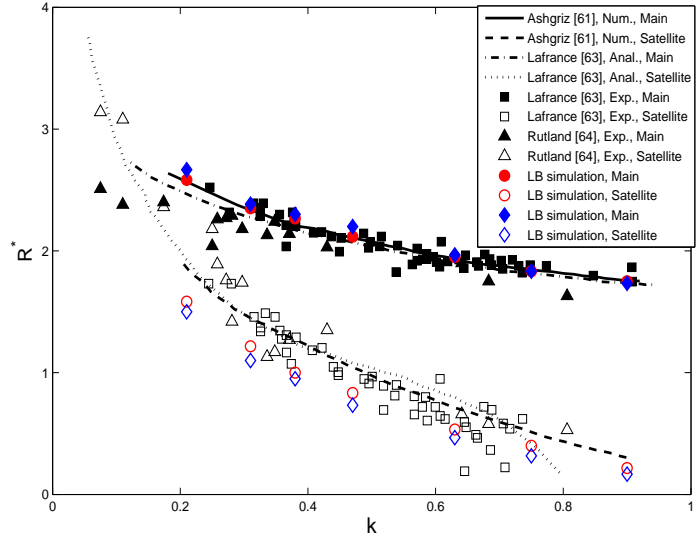}
 \tiny\caption{The terminal droplet sizes at different wave numbers. Here the red circle and blue rhombus respectively represent the cases of $\rho_l/\rho_g=10$ and $\rho_l/\rho_g=100$, and the droplet radius $R^*$ has been normalized by $R$.  }
\end{figure}


\section{Summary}\label{sec:sum}

Numerical modeling of axisymmetric multiphase flows with large
density contrasts is still an intractable task in the framework of
the LB method. In this paper, we propose a simple and efficient LB
model for axisymmetric multiphase system, which is capable of
handling large density differences. The proposed LB model is built
upon the conservative phase-field equation, which involves a
lower-order diffusion term as opposed to the widely used
Cahn-Hilliard equation. Therefore the present model in theory can
achieve better numerical stability and accuracy in solving phase
interface than the Cahn-Hilliard type of axisymmetric LB models. Two
LB evolution equations are utilized in the current model, one of
which is used for capturing phase interface and the other for
solving fluid velocity and pressure. A new equilibrium distribution
function and some discrete source terms are designed in the LB
equation for interface capturing. Meanwhile, a simpler forcing
distribution function is also proposed in the LB equation for
hydrodynamics. Different from most of previous axisymmetric LB
multiphase models~\cite{Premnath, Huang, Mukherjee, HuangJJ,
Srivastava}, the added source terms accounting for the axisymmetric
effect contain no gradients in the present model. The present model
is also equipped with an advanced MRT collision operator to enhance
its numerical stability. We conducted the Chapman-Enskog analysis on
the present MRT-LB equations, and it is demonstrated that both the
conservative AC equation and the incompressible NS equations in the
cylindrical coordinate system can be recovered correctly from the
present model. To validate the present model, we first simulated a
steady problem of the static droplet with a high density ratio of
1000, and it is found that the present model can accurately solve
the density field, and also achieve low spurious velocities with the
order of $10^{-6}$. To further assess the current model, two dynamic
benchmark problems of a droplet oscillation and a liquid thread
breakup are considered. It is shown that the present model can
provide satisfactory predictions of the droplet oscillation
frequency and the generated daughter droplet sizes for a board range
of density ratios. At last, we simulated the realistic
buoyancy-driven bubbly flow with a large density ratio of 1000. Some
fascinating bubble dynamics are successfully reproduced by the
present model, and the numerically predicted terminal bubble
patterns show good agreements with the experimental date and some
available numerical results. It is also reported that the present
model can describe bubble interfacial dynamics with a higher
accuracy than the existing axisymmetric LB model~\cite{Huang}. The present
method is developed based on the standard orthogonal MRT model, and its extension
to the non-orthogonal MRT one can be conducted directly. It is expected that the non-orthogonal
model can retain the numerical accuracy while simplifying the implementation~\cite{Rosis1, Rosis2}. In conclusion, we anticipate
that our present numerical method will be useful for many practical and sophisticated problems.

\section*{Acknowledgments}
This work is also financially supported by the National Natural
Science Foundation of China under Grants Nos. (11602075, 51776068 and 11674080),
and Natural Science Foundation of Zhejiang Province (No. LR17A050001).

\appendix
\section{Chapman-Enskog analysis of axisymmetric LB model for the Allen-Cahn equation}
The Chapman-Enskog analysis is now carried out to demonstrate the
consistency of the LB evolution equation (9) with the axisymmetric
AC equation. We first introduce the multiscaling expansions for the
distribution function, time derivative, space gradient, and discrete
source term as
\begin{subequations}
\begin{equation}
{{f}_i} = {f}_i^{(0)} + \epsilon {f}_i^{(1)} + {\epsilon
^2}{f}_i^{(2)} + \cdot  \cdot  \cdot ,
\end{equation}
\begin{equation}
{\partial _t} = \epsilon {\partial _{t_1}} + {\epsilon ^2}{\partial
_{t_2}},\nabla  = \epsilon {\nabla _1}, F_i = \epsilon
{F_i}^{(1)}+\epsilon^2 {F_i}^{(2)},
\end{equation}
\end{subequations}
where $\epsilon$ is a small expansion parameter. Applying Taylor
expansion to Eq. (9) about $\textbf{x}$ and $t$, one can derive the
resulting continuous equation
\begin{equation}
{D_i}{{f}_i} + {{{\delta _t}} \over 2}D_i^2{{f}_i} +  \cdot \cdot
\cdot =  - {1 \over {{\delta _t}}}{({{\bf{M}}^{ -
1}}{{\bf{S}}^f}{\bf{M}})_{ij}}({{f}_j} - {f}_j^{eq}) +{F_i},
\end{equation}
where $D_i=\partial_t+\textbf{c}_i\cdot\nabla$. Substituting the
expansions into Eq. (A.2), we can obtain the following equations in
consecutive order of the parameter $\epsilon$,
\begin{subequations}
\begin{equation}
{\epsilon^0}:\qquad f_i^{(0)} = f_i^{(eq)},
\end{equation}
\begin{equation}
{\epsilon ^1}:\qquad {D_{1i}}f_i^{(0)} =  - {1 \over {{\delta
_t}}}{({{\bf{M}}^{ - 1}}{{\bf{S}}^f}{\bf{M}})_{ij}}f_j^{(1)} +
{\left[ {{{\bf{M}}^{ - 1}}({\rm \textbf{I}} - {{{{\bf{S}}^f}} \over
2}){\bf{M}}} \right]_{ij}}{{R_j}}^{(1)},
\end{equation}
\begin{equation}
{\epsilon ^2}:\qquad {\partial _{{t_2}}}f_i^{(0)} +
{D_{1i}}f_i^{(1)} + {{{\delta _t}} \over 2}D_{1i}^2f_i^{(0)} =  - {1
\over {{\delta
_t}}}{({{\bf{M}}^{-1}}{{\bf{S}}^f}{\bf{M}})_{ij}}f_j^{(2)} + {\left[
{{{\bf{M}}^{ - 1}}({\rm \textbf{I}} - {{{{\bf{S}}^f}} \over
2}){\bf{M}}} \right]_{ij}}{R_j}^{(2)},
\end{equation}
\end{subequations}
in which ${D_{1i}} = {\partial _{{t_1}}} + {{\bf{c}}_i} \cdot
{\nabla _1}$. Rewriting Eqs. (A.3) in vector form and premultiplying
the matrix $\textbf{M}$ on both sides of them, we can derive the
corresponding equations in the moment space
\begin{subequations}
\begin{equation}
{\epsilon ^0}:\qquad {{\bf{m}}_f^{(0)}} = {{\bf{m}}_f^{(eq)}},
\end{equation}
\begin{equation}
{\epsilon ^1}:\qquad {{\hat{\bf{D}}}_1}{{\bf{m}}_f^{(0)}} =  -
{{\bf{S}}^{f'}}{{\bf{{m}}}_f^{(1)}} + ({\rm \textbf{I}} -
{{{{\bf{S}}^f}} \over 2}){\bf{M}}{{\bf{{R}}}^{(1)}},
\end{equation}
\begin{equation}
{\epsilon^2}:\qquad {\partial _{{t_2}}}{{\bf{m}}_f^{(0)}} +
{\hat{{\bf{D}}}_1}({\rm \textbf{I}} - {{\bf{S}}^f \over
2}){{\bf{{m}}}_f^{(1)}} + {{{\delta _t}} \over
2}{\hat{\textbf{D}}_1}({\rm \textbf{I}} - {{\bf{S}}^f \over
2}){\bf{M}}{{{\bf{R}}}^{(1)}} =
-{{\bf{S}}^{f'}}{{\bf{{m}}}_f^{(2)}}+({\rm \textbf{I}} -
{{{{\bf{S}}^f}} \over 2}){\bf{M}}{{\bf{{R}}}^{(2)}},
\end{equation}
\end{subequations}
where ${{\bf{S}}^{f'}}={{\bf{S}}^{f}}/\delta_t$,
${\hat{{\bf{D}}}_1}=\textbf{M}{\textbf{D}_1}\textbf{M}^{-1}$,
${{\bf{D}}_1} = {\partial _{{t_1}}}{\rm \textbf{I}} + {\nabla _1}
\cdot diag({{\bf{c}}_0},{{\bf{c}}_1}, \cdot  \cdot  \cdot
,{{\bf{c}}_8})$, ${\bf{{m}}}_f^{(1)}=(m_{f_0}^{(1)},m_{f_1}^{(1)},
\cdot  \cdot  \cdot,m_{f_8}^{(1)})$ and Eq. (A3.b) has been used to
derive the resulting equation (A.4c). Substituting Eq. (21) to Eq.
(A.4b), we can obtain several equations related to the target
governing equations,
\begin{subequations}
\begin{equation}
{\partial _{{t_1}}}(r\phi)  + {\partial _{{z_1}}}(r\phi {u_z}) +
{\partial _{{r_1}}}(r\phi {u_r}+M\phi) = 0 ,
\end{equation}
\begin{equation}
{\partial _{{t_1}}}(r\phi {u_z}) + {\partial
_{{z_1}}}({c_s^2}r\phi)=-c{s_3^{f'}}m_{f_3}^{(1)}+c(1-\frac{{s_3^f}}{2}){mR_1^{(1)}},
\end{equation}
\begin{equation}
{\partial _{{t_1}}}(r\phi {u_r}+M\phi) + {\partial
_{{r_1}}}({c_s^2}r\phi)=-c{s_5^{f'}}m_{f_5}^{(1)}+c(1-\frac{{s_5^f}}{2}){mR_2^{(1)}}.
\end{equation}
\end{subequations}
Similarly, the substitution of Eq. (21) into Eq. (A.4c) yields
\begin{equation}
{\partial _{{t_2}}}(r\phi) +{\partial
_{{z_1}}}{c(1-\frac{{s_3^f}}{2})m_{f_3}^{(1)}}
+{\partial_{{r_1}}}c(1-\frac{{s_5^f}}{2})m_{f_5}^{(1)} +
\frac{\delta_t}{2}\partial_{z_{1}}c(1-\frac{{s_3^f}}{2}){mR_1^{(1)}}+
\frac{\delta_t}{2}\partial_{r_{1}}c(1-\frac{{s_5^f}}{2}){mR_2^{(1)}}=0,
\end{equation}
where the unknown $m_{f_3}^{(1)}$ and $m_{f_5}^{(1)}$ can be
determined from Eqs. (A.5b-A.5c), and then their expressions can be
given by
\begin{subequations}
\begin{equation}
c{s_3^{f'}}m_{f_3}^{(1)}=-{\partial _{{t_1}}}(r\phi {u_z})-{\partial
_{{z_1}}}({c_s^2}r\phi)+c(1-\frac{{s_3^f}}{2}){mR_1^{(1)}},
\end{equation}
\begin{equation}
c{s_5^{f'}}m_{f_5}^{(1)}=-{\partial _{{t_1}}}(r\phi {u_r}+M\phi) -
{\partial_{{r_1}}}({c_s^2}r\phi)+c(1-\frac{{s_5^f}}{2}){mR_2^{(1)}}.
\end{equation}
\end{subequations}
Substituting Eqs. (A7) to Eq. (A6), we have
\begin{equation}
\partial_{t_2}(r\phi)=\partial_{z_1}c_s^2 \delta_t( \frac{1}{s_3^f}-\frac{1}{2} )\left[\partial_{z_1}(r\phi)-r\lambda n_z^{(1)}\right]
+\partial_{r_1}c_s^2
\delta_t(\frac{1}{s_5^f}-\frac{1}{2})\left[\partial_{r_1}(r\phi)-r\lambda n_r^{(1)}\right],
\end{equation}
where Eq. (23) has been applied. Combining Eq. (A.5a) at $t_1$ time
scale and Eq. (A.8) at $t_2$ time scale, we can derive the recovered
equation,
\begin{equation}
\partial_t(r\phi)+\partial_\alpha(r\phi u_\alpha+M\phi\delta_{\alpha r})=\partial_\alpha(M\partial_\alpha r\phi)-M\partial_\alpha(r\lambda
n_\alpha),
\end{equation}
where the relation $s_3^f=s_5^f$ has been used to obtain the
isotropic mobility,
\begin{equation}
 M= c_s^2 \delta_t( \frac{1}{s_3^f}-\frac{1}{2} ).
\end{equation}
From the above procedure, it is clear that the axisymmetric AC
equation can be recovered correctly from the present MRT model
without adopting any approximations.

\section{Chapman-Enskog analysis of axisymmetric LB model for the incompressible hydrodynamic equations}

The present MRT-LB model for the axisymmetric NS equations is also
analyzed by applying the Chapman-Enskog expansions, and similarly
the particle distribution function, time and space derivatives,
forcing distribution function can be expanded as
\begin{subequations}
\begin{equation}
{{g}_i} = {g}_i^{(0)} + \epsilon {g}_i^{(1)} + {\epsilon
^2}{g}_i^{(2)} + \cdot  \cdot  \cdot ,
\end{equation}
\begin{equation}
{\partial _t} = \epsilon {\partial _{t_1}} + {\epsilon ^2}{\partial
_{t_2}},\nabla  = \epsilon {\nabla _1}, G_i =\epsilon {G_i}^{(1)}.
\end{equation}
\end{subequations}
Adopting Taylor expansion to Eq. (24) and substituting the
expansions (B.1) into the resulting equation, one can get the zero-,
first-, and second-order equations in the parameter $\epsilon$,
\begin{subequations}
\begin{equation}
{\epsilon^0}:\qquad {g}_i^{(0)} = {g}_i^{(eq)},
\end{equation}
\begin{equation}
{\epsilon ^1}:\qquad {D_{1i}}{g}_i^{(0)} = - {1 \over {{\delta
_t}}}{({{\bf{M}}^{ - 1}}{{\bf{S}}^g}{\bf{M}})_{ij}}{g}_j^{(1)}
+{{G_i}}^{(1)},
\end{equation}
\begin{equation}
{\epsilon ^2}:\qquad {\partial _{{t_2}}}{g}_i^{(0)} +
{D_{1i}}{g}_i^{(1)} + {{{\delta _t}} \over 2}D_{1i}^2{g}_i^{(0)} = -
{1 \over {{\delta _t}}}{({{\bf{M}}^{ -
1}}{{\bf{S}}^g}{\bf{M}})_{ij}}{g}_j^{(2)}.
\end{equation}
\end{subequations}
Substituting Eq. (B.2b) into Eq. (B.2c) and multiplying matrix
$\textbf{M}$ on the both sides of Eqs. (B2) separately gives
\begin{subequations}
\begin{equation}
{\epsilon ^0}:\qquad {{\bf{m}}_g^{(0)}} = {{\bf{m}}_g^{(eq)}},
\end{equation}
\begin{equation}
{\epsilon ^1}:\qquad {{\hat{\bf{D}}}_1}{{\bf{m}}_g^{(0)}} =  -
{{\bf{S}}^{g'}}{{\bf{{m}}}_g^{(1)}} + ({\rm \textbf{I}} -
{{{{\bf{S}}^g}} \over 2}){\bf{M}}{{\bf{{T}}}^{(1)}},
\end{equation}
\begin{equation}
{\epsilon^2}:\qquad {\partial _{{t_2}}}{{\bf{m}}_g^{(0)}} +
{\hat{{\bf{D}}}_1}({\rm \textbf{I}} - {{\bf{S}}^g \over
2}){{\bf{{m}}}_g^{(1)}} + {{{\delta _t}} \over
2}{\hat{\textbf{D}}_1}({\rm \textbf{I}} - {{\bf{S}}^g \over
2}){\bf{M}}{{{\bf{T}}}^{(1)}} =
-{{\bf{S}}^{g'}}{{\bf{{m}}}_g^{(2)}},
\end{equation}
\end{subequations}
where ${{\bf{S}}^{g'}}={{\bf{S}}^{g}}/\delta_t$. Now it is seen that
the distribution functions in the discrete-velocity space can be
projected onto the macroscopic quantities in the moment space. Based
on the moment conditions, we can define ${{\bf{{m}}}_g}$ as
\begin{equation}
{{\bf{{m}}}_g}={\bf{M}}{{\bf{g}}}=( - {{{\delta _t}} \over
2}{r}{\bf{u}} \cdot \nabla \rho ,{{m}_{g_1}},{{m}_{g_2}},{{r\rho
{u_z}} \over c}-{{{\delta _t}{F_{z}} } \over
{2c}},{{m}_{g_4}},{{r\rho {u_r}} \over c} - {{{\delta_t}{F_{r}}}
\over {2c}},{{m}_{g_6}},{{m}_{g_7}},{{m}_{g_8}})^{\rm T},
\end{equation}
and ${{\bf{{m}}}_g^{(k)}}(k\geq1)$ can be denoted by
\begin{subequations}
\begin{equation}
{{\bf{{m}}}_g^{(1)}} = ( - {{{\delta _t}} \over 2}{r\bf{u}} \cdot
{\nabla _1}\rho ,{m}_{g_1}^{(1)},{m}_{g_2}^{(1)}, - {{{\delta
_t}{F_z}^{(1)}} \over {2c}},{m}_{g_4}^{(1)}, - {{{\delta
_t}{F_r}^{(1)}\over {2c}},{ m}_{g_6}^{(1)}}
,{m}_{g_7}^{(1)},{m}_{g_8}^{(1)})^{\rm T},
\end{equation}
\begin{equation}
{{\bf{{m}}}_g^{(k)}}=(0,{m}_{g_1}^{(k)},{m}_{g_2}^{(k)},0,{m}_{g_4}^{(k)},0,{m}_{g_6}^{(k)},{m}_{g_7}^{(k)},{m}_{g_8}^{(k)})^{\rm
T}, \qquad (k\geq2).
\end{equation}
\end{subequations}
Substituting Eqs. (36-37) and Eq. (B.5a) into Eq. (B.3b), we can
obtain several macroscopic equations at the $\epsilon$ scale, and
only ones used to recover the hydrodynamic equations are presented
here,
\begin{subequations}
\begin{equation}
{\partial _{{z_1}}}{(ru_z)} + {\partial _{{r_1}}}{(ru_r)} = 0 ,
\end{equation}
\begin{equation}
{\partial_{{t_1}}}(r\rho {u_z}) + {\partial _{{z_1}}}(rp + r\rho
u_z^2) + {\partial _{{r_1}}}(r\rho {u_z}{u_r}) = F_{z}^{(1)},
\end{equation}
\begin{equation}
{\partial _{{t_1}}}(r\rho {u_r}) + {\partial_{{z_1}}}r\rho
{u_z}{u_r} + {\partial_{{r_1}}}(rp + r\rho u_r^2) = F_{r}^{(1)},
\end{equation}
\begin{equation}
{\partial _{{t_1}}}{{2rp + r\rho u_z^2 + r\rho u_r^2} \over {c_s^2}}
= -s_1^{g'}{m}_{{g_1}}^{(1)} + {T_{{g_1}}},
\end{equation}
\begin{equation}
{\partial _{{t_1}}}{{r\rho u_z^2 - r\rho u_r^2} \over {{c^2}}} +
{\partial_{{z_1}}}{{2r\rho {u_z}} \over 3} -
{\partial_{{r_1}}}{{2r\rho {u_r}} \over 3} =  -
s_7^{g'}{m}_{{g_7}}^{(1)} + {T_{{g_7}}},
\end{equation}
\begin{equation}
{\partial _{{t_1}}}{{r\rho {u_z}{u_r}} \over {{c^2}}} +
{\partial_{{z_1}}}{{r\rho {u_r}} \over 3} + {\partial
_{{r_1}}}{{r\rho {u_z}} \over 3} =- s_8^{g'}{m}_{{g_8}}^{(1)} +
{T_{{g_8}}},
\end{equation}
\end{subequations}
where
\begin{subequations}
\begin{equation}
{T_{{g_1}}} =2(1-{{s_1^g} \over 2})(\rho{u_r}),
\end{equation}
\begin{equation}
{T_{{g_7}}} = {{2\over{{3}}}(1-{{s_7^g} \over 2})\left[
{{u_z}{\partial_{z_1}}(r\rho)-{u_r}{\partial_{r_1}}(r\rho)}\right]},
\end{equation}
\begin{equation}
{T_{{g_8}}} ={ { {1\over {{3}}}(1-{{s_8^g} \over 2})\left[
{{u_z}{\partial_{r_1}}(r\rho) +
{u_r}{\partial_{z_1}}(r\rho)}\right]}}.
\end{equation}
\end{subequations}
Similarly, we substitute Eqs. (36-37) and Eqs. (B.5) into Eq.
(B.3c). With some algebraic operations, the moment equations at the
$\epsilon^2$ scale can be derived and the related ones are presented
as
\begin{subequations}
\begin{equation}
{\partial_{{t_2}}}(r\rho {u_z}) + {{{c^2}} \over
6}{\partial_{{z_1}}}(1 - {{s_1^g} \over 2}){m}_{g_1}^{(1)} + {{{c^2}}
\over 2}{\partial _{{z_1}}}(1 - {{s_7^g} \over 2}){m}_{g_7}^{(1)} +
{c^2}{\partial _{{r_1}}}(1 - {{s_8^g} \over 2}){m}_{g_8}^{(1)} +
{T_{gz178}} = 0,
\end{equation}
\begin{equation}
{\partial_{{t_2}}}(r\rho {u_r}) + {{{c^2}} \over 6}{\partial
_{{r_1}}}(1 - {{s_1^g} \over 2}){m}_{g_1}^{(1)} - {{{c^2}} \over
2}{\partial _{{r_1}}}(1 - {{s_7^g} \over 2}){m}_{g_7}^{(1)} +
{c^2}{\partial_{{z_1}}}(1 - {{s_8^g} \over 2}){m}_{g_8}^{(1)} +
{T_{gr178}} = 0,
\end{equation}
\end{subequations}
where ${T_{gz178}}$ and ${T_{gr178}}$ are given by
\begin{eqnarray}
{T_{gz178}} ={\partial_{{z_1}}}(1 - {{s_1^g} \over 2}){
{{{c^2}{\delta_t}} \over 6}{(\rho{u_r})} } + {\partial _{{z_1}}}(1
- {{s_7^g} \over 2}){{{c^2}{\delta_t}}\over 6} { \left[{{u_z}{\partial_{z_1}}(r\rho ) -{u_r}{\partial_{r_1}}(r\rho )}\right]} \nonumber\\
+ {\partial _{{r_1}}}(1-{{s_8^g}\over 2}){{
{c^2{\delta_t}\over{{6}}}[ {{u_z}{\partial_{r_1}}(r\rho)
+{u_r}{\partial_{z_1}}(r\rho )}]}},
\end{eqnarray}
and
\begin{eqnarray}
{T_{gr178}} = {\partial_{{r_1}}}(1 - {{s_1^g} \over 2}){
{{{c^2}{\delta_t}} \over 6}{(\rho{u_r})}}-{\partial
_{{r_1}}}(1-{{s_7^g} \over 2}){{{c^2}{\delta_t}}\over 6}{[{{u_z}{\partial_{z_1}}(r\rho) -{u_r}{\partial_{r_1}}(r\rho)}]} \nonumber\\
+{\partial_{{z_1}}}(1-{{s_8^g} \over
2}){{{{c^2}{\delta_t}\over{{6}}}[{{u_z}{\partial_{r_1}}(r\rho )
+{u_r}{\partial_{z_1}}(r\rho)}]}},
\end{eqnarray}
In Eqs. (B.8), the variables ${m}_{g_1}^{(1)}$, ${m}_{g_7}^{(1)}$ and
${m}_{g_8}^{(1)}$ needs to be determined. Based on the relations
(B.6), one can get
\begin{subequations}
\begin{equation}
- s_1^{g'}{m}_{{g_1}}^{(1)}={\partial_{{t_1}}}{{2rp+r\rho u_z^2+r\rho u_r^2} \over {c_s^2}} - 2(1-{{s_1^{g}} \over 2})\rho{u_r},
\end{equation}
\begin{equation}
- s_7^{g'}{m}_{{g_7}}^{(1)}={\partial_{{t_1}}}{{r\rho u_z^2-r\rho u_r^2} \over {c^2}} + {{s_7^g} \over {{3}}}\left[ {{u_z}{\partial_{z_1}}(r\rho) - {u_r}{\partial_{r_1}}(r\rho)} \right] +{2 \over 3} r \rho
\left[{\partial_{{z_1}}}{u_z} - {\partial_{{r_1}}}{u_r}\right],
\end{equation}
\begin{equation}
 - s_8^{g'}{m}_{{g_8}}^{(1)}={\partial_{{t_1}}}{{r\rho u_z u_r} \over {c^2}} + {{s_8^g} \over {{6}}}\left[ {{u_z}{\partial_{r_1}}(r\rho) + {u_r}{\partial_{z_1}}(r\rho)} \right] +{1 \over 3} r \rho
\left[{\partial_{{r_1}}}{u_z} + {\partial_{{z_1}}}{u_r}\right].
\end{equation}
\end{subequations}
Substituting Eqs. (B.11) into Eqs. (B.8) and using the incompressible limits [$\partial_{t_1} p=O(Ma^2)$ and $\mathbf{u}^2=O(Ma^2)$], we can ultimately derive the second-order equations in $\epsilon$,
\begin{subequations}
\begin{equation}
{\partial _{{t_2}}}(r\rho {u_z}) - {\partial_{{z_1}}}r\nu\rho
({\partial_{{z_1}}}{u_z} - {\partial_{{r_1}}}{u_r}-\frac{u_r}{r}) - {\partial
_{{r_1}}}r\nu\rho ( {\partial _{{r_1}}}{u_z}+
{\partial_{{z_1}}}{u_r}) = 0,
\end{equation}
\begin{equation}
{\partial _{{t_2}}}(r\rho {u_r})-{\partial_{{z_1}}}r\nu\rho
({\partial_{{z_1}}}{u_r} + {\partial_{{r_1}}}{u_z}) - {\partial
_{{r_1}}}r\nu\rho ({\partial _{{r_1}}}{u_r}-{\partial_{{z_1}}}{u_z}-\frac{u_r}{r}) = 0,
\end{equation}
\end{subequations}
where the terms with the order of $O(\delta_t Ma^2)$ have been neglected, and the kinematic viscosity can be given by
\begin{equation}
\nu=c_s^2 \delta_t (\tau_g-{1 \over 2}),
\end{equation}
where $1/\tau_g=s_1^g=s_7^g=s_8^g$. Substituting Eq. (B.6a) into Eqs. (B.12) and
further combining Eqs.(B.6a-B.6c) and Eqs. (B.12a-B.12b) at $t_1$ and $t_2$
time scales, we can obtain the hydrodynamic equations,
\begin{subequations}
\begin{equation}
\partial_\alpha(r u_\alpha)=0,
\end{equation}
\begin{equation}
\partial_t (r\rho {u_\beta})+{\partial_\alpha}(r\rho
{u_\alpha}u_\beta)=-\partial_\beta(rp)+{\partial_\alpha}[r\nu\rho({\partial_\alpha}u_\beta+{\partial_\beta}u_\alpha)]+{F_\beta}.
\end{equation}
\end{subequations}
From the above discussions, it is shown that our MRT-LB model can
exactly recover the axisymmetric hydrodynamic equations under the
incompressible limits.

\section{Calculation of the hydrodynamic pressure}
Lastly, we give a discussion on the calculation of the hydrodynamic
pressure $p$. Based on Eq. (25), we have
\begin{equation}
g_0^{(eq)}({\bf{x}},t) = {{r({\omega _0} - 1)} \over
{c_s^2}}p({\bf{x}},t) + r\rho {s_0}({\bf{u}}({\bf{x}},t)) ,
\end{equation}
which can be further recast as
\begin{equation}
g_0({\bf{x}},t)-[g_0({\bf{x}},t)-g_{0}^{(eq)}({\bf{x}},t)] =
{{r({\omega _0} - 1)} \over {c_s^2}}p({\bf{x}},t) + r\rho
{s_0}({\bf{u}}({\bf{x}},t)).
\end{equation}
Expanding $g_i(\textbf{x}+\textbf{c}_i\delta_t,t+\delta_t)$ in Eq.
(24) about $\bf{x}$ and $t$, we can easily derive
\begin{equation}
{\delta_t}D_i{g_i}(\textbf{x},t)=-\left[{{\bf{M}}^{
-1}}{{\bf{S}}^g}{\bf{M}}\right]_{ij}\left[g_j(\textbf{x},t)-g_j^{eq}(\textbf{x},t)\right]
+\delta_t{G}_i(\textbf{x},t)
\end{equation}
to the order of ${O}(\delta _t^2)$. Premultiplying the matrix
$\textbf{M}^{-1}{\textbf{S}^g}^{-1}\textbf{M}$ on both sides of Eq.
(C.3), we can get
\begin{equation}
{g_i}({\bf{x}},t) - g_i^{(eq)}({\bf{x}},t) =
-{\delta_t}{({{\bf{M}}^{ - 1}}{{\bf{S}}^g}^{-1}{\bf{M}})_{ij}}{D_j}{
g_j}({\bf{x}},t) + {\delta_t}{({{\bf{M}}^{ -
1}}{{\bf{S}}^g}^{-1}{\bf{M}})_{ij}}{G_j}({\bf{x}},t),
\end{equation}
which indicates
\begin{equation}
g_i({\bf{x}},t) = g_i^{(eq)}({\bf{x}},t) + O({\delta _t}).
\end{equation}
With the above approximate, we can rewrite Eq. (C.4) as
\begin{equation}
{g_i}({\bf{x}},t) - g_i^{(eq)}({\bf{x}},t) =  - {\delta
_t}{({{\bf{M}}^{ - 1}}{{\bf{S}}^g}^{ - 1}{\bf{M}})_{ij}}{D_j}{
g_j^{(eq)}}({\bf{x}},t) + {\delta _t}{({{\bf{M}}^{ -
1}}{{\bf{S}}^g}^{-1}{\bf{M}})_{ij}}{G_j}({\bf{x}},t).
\end{equation}
Substituting Eqs. (36) and (37) into Eq. (C.6), and taking $i=0$, we
can have
\begin{equation}
{g_0}({\bf{x}},t)-g_0^{(eq)}({\bf{x}},t) =({{3{s_1^g} + 2{s_2^g}}
\over {3{c^2}{s_1^g}{s_2^g}}}){\delta_t}{{\partial rp} \over
{\partial t}} + ({{{s_1^g} + {s_2^g}} \over
{3{c^2}{s_1^g}{s_2^g}}}){\delta_t}{{\partial
(r\rho{\bf{u}}\cdot{\bf{u}})} \over {\partial t}} -{{{s_1^g} +
{s_2^g} -{s_1^g}{s_2^g}} \over {2{s_1^g}{s_2^g}}}\cdot
{\delta_t}\omega_0 \rho u_r
\end{equation}
to the order of ${O}(\delta _t^2)$. Under the incompressible
condition, $\partial_t p=O(Ma^2)$ and $|\textbf{u}|=O(Ma)$ are
satisfied, and then we can rewrite Eq. (C.7) as
\begin{equation}
{g_0}({\bf{x}},t) - g_0^{(eq)}({\bf{x}},t) = -{{{s_1^g} + {s_2^g}
-{s_1^g}{s_2^g}} \over {2{s_1^g}{s_2^g}}}\cdot {\delta_t}\omega_0
\rho u_r + O({{\delta _t}^2} + \delta _t Ma^2).
\end{equation}
Substituting Eq. (C.8) into Eq. (C.2) yields
\begin{equation}
{r({{\omega _0} - 1}) \over {c_s^2}}p ={{g_0}}-r\rho{s_0}({\bf{u}})
+{{{s_1^g} + {s_2^g} -{s_1^g}{s_2^g}} \over {2{s_1^g}{s_2^g}}}\cdot
{\delta_t}\omega_0 \rho u_r
\end{equation}
to the order of $O(\delta_t^2+{\delta_t}M{a^2})$. Considering the
first component in Eq. (B.4), the zeroth-order moment of the
distribution function can be defined by
\begin{equation}
\sum\limits_{i}{g_i}=-\frac{\delta_t}{2}{r}{\bf{u}} \cdot \nabla
\rho,
\end{equation}
and the hydrodynamic pressure $p$ can be evaluated as
\begin{equation}
p=\frac{c_s^2}{1-\omega_0}\left[\frac{1}{r}\sum_{i\neq0}g_i+\frac{\delta_t}{2}\textbf{u}\cdot\nabla\rho+\rho
s_0(\textbf{u})- {{{s_1^g} + {s_2^g} -{s_1^g}{s_2^g}} \over
{2{s_1^g}{s_2^g}}}\cdot \frac{{\delta_t}\omega_0\rho u_r}{r}\right].
\end{equation}


\clearpage

\end {document}